\documentclass[
 reprint,
 amsmath,
 pre,
 amssymb,
 nofootinbib,
 aps,
]{revtex4-1}

\usepackage[pdftex]{graphicx}
\usepackage{wasysym}
\usepackage{amsfonts}
\usepackage{ifsym}

\makeatletter
\newlength{\apb@width}
\newcommand{\autoparbox}[2][c]{\settowidth{\apb@width}{#2}\parbox[#1]{\apb@width}{#2}}

\makeatother

\newcommand{\namedref}[2]{\hyperref[#2]{#1~\ref*{#2}}}

\newcommand{\eps}{\varepsilon}

\renewcommand{\Re}{\mathop{\mathrm{Re}}}
\renewcommand{\Im}{\mathop{\mathrm{Im}}}

\newcommand{\Csphere}{{}^\bullet\kern-1.2pt C}
\newcommand{\Ctorus}{{}^\circ\kern-1.2pt C}

\newcommand{\nn}{\nonumber}

\newcommand{\COMMENT}[1]{}

\newcommand{\neqa}{\nonumber\end{eqnarray}}
\newcommand{\la}[1]{\label{#1}}

\newcommand{\rf}[1]{(\ref{#1})}

\newcommand{\<}{{\langle}}
\renewcommand{\>}{{\rangle}}

\newcommand{\re}{\relax{\rm I\kern-.18em R}}

\def\su2{{SU(2)}}

\def\eps{{\epsilon}}

\def\a{{\alpha}}

\def\[{\left[}
\def\]{\right]}

\def\om{\omega}

\def\a{\alpha}

\def\({\left(}
\def\){\right)}
\def\[{\left[}
\def\]{\right]}

\def\<{\langle}
\def\>{\rangle}

\def\i2{\frac{i}{2}}

\def\2F1{\,_2{\rm F}_1}

\usepackage[usenames,dvipsnames]{xcolor}
\usepackage{color}
\usepackage{graphicx}
\usepackage{dcolumn}
\usepackage{bm}
\usepackage{hyperref}

\usepackage{cancel}
\usepackage{ctable}
\usepackage{booktabs}
\newcolumntype{L}[1]{>{\raggedright\let\newline\\\arraybackslash\hspace{0pt}}m{#1}}
\newcolumntype{C}[1]{>{\centering\let\newline\\\arraybackslash\hspace{0pt}}m{#1}}
\newcolumntype{R}[1]{>{\raggedleft\let\newline\\\arraybackslash\hspace{0pt}}m{#1}}

\newcommand{\beq}{\begin{equation}}
\newcommand{\eeq}{\end{equation}}
\newcommand{\beqq}{\begin{equation*}}
\newcommand{\eeqq}{\end{equation*}}
\newcommand\beqa{\begin{eqnarray}}
\newcommand\eeqa{\end{eqnarray}}
\newcommand\beqaa{\begin{eqnarray*}}
\newcommand\eeqaa{\end{eqnarray*}}
\newcommand\bea{\begin{array}}
\newcommand\eea{\end{array}}

\begin{document}


\title{Flux Tube S-matrix Bootstrap}

\author{Joan Elias Mir\'o}
\affiliation{CERN, Theoretical Physics Department, \\
Rte de Meyrin 385, CH-1211, Geneva, Switzerland}
\author{Andrea L. Guerrieri}
\affiliation{Instituto de F\'isica Te\'orica, UNESP, ICTP South American Institute for Fundamental Research, Rua Dr Bento Teobaldo Ferraz 271, 01140-070, S\~ao Paulo, Brazil}
\author{Aditya Hebbar}
\affiliation{Fields and String Laboratory, Institute of Physics, \'Ecole Polytechnique F\'ed\'erale de Lausanne (EPFL), \\
Rte de la Sorge, BSP 728, CH-1015 Lausanne, Switzerland}
\author{Jo\~ao Penedones}
\affiliation{Fields and String Laboratory, Institute of Physics, \'Ecole Polytechnique F\'ed\'erale de Lausanne (EPFL), \\
Rte de la Sorge, BSP 728, CH-1015 Lausanne, Switzerland}
\author{Pedro Vieira}
\affiliation{Perimeter Institute for Theoretical Physics, 31 Caroline St N Waterloo, Ontario N2L 2Y5, Canada}
\affiliation{Instituto de F\'isica Te\'orica, UNESP, ICTP South American Institute for Fundamental Research, Rua Dr Bento Teobaldo Ferraz 271, 01140-070, S\~ao Paulo, Brazil}


\begin{abstract}
We bootstrap the S-matrix of massless particles in unitary, relativistic two dimensional quantum field theories.
We find that the low energy expansion of such S-matrices is strongly constrained by the existence of a UV completion. In the context of flux tube physics, this allows us to constrain several terms in the   
S-matrix low energy expansion or -- equivalently -- on Wilson coefficients of several irrelevant operators showing up in the flux tube effective action. These bounds have direct implications for other physical quantities; for instance, they allow us to further bound the ground state energy 
as well as the level splitting of degenerate energy levels of large flux tubes. 
We find that the S-matrices living at the boundary of the allowed space
exhibit an intricate pattern of  resonances  with one sharper resonance whose quantum numbers, mass and width are precisely those of the world-sheet axion 
proposed in 
\cite{Athenodorou:2010cs, Dubovsky:2013gi}. The general method proposed here should be extendable to 
massless S-matrices in higher dimensions and should lead to 
new quantitative bounds on irrelevant operators in theories of Goldstones and also in gauge and gravity theories.
\end{abstract}

\pacs{Valid PACS appear here}
\maketitle

\section{Introduction}
\label{sec:introduction}

Unraveling the dual string description of  Yang-Mills theory is an old-standing problem. A first step towards achieving this goal is solving for the spectrum of long strings or confining  flux tubes of pure glue.
At low energies, the massless flux tube excitations (or branons) decouple from the massive short strings (or glueballs)\footnote{If the number of colours $N_c$ tends to infinity, then the flux tubes decouple from the   glueballs  at any   energy (independent of $N_c$). 
}
and can be described by a two dimensional worldsheet theory which 
can be formulated in terms of an effective Lagrangian or in terms of the branon S-matrix. 
Both approaches have their advantages and limitations. 

The flux tube's effective Lagrangian density is built out of derivatives of the fields $X^\mu$ describing the embedding of the worldsheet in spacetime. At low energies, it is dominated by the square root of the  induced metric determinant $h=\text{det}\,h_{\alpha\beta} =\text{det}\, \partial_\alpha X^\mu \partial_\beta X^\nu\eta_{\mu\nu}$, i.e.\ the Nambu-Goto lagrangian.  Any  interaction consistent with the  bulk $D$-dimensional Poincar\'e symmetry is also  permitted. Thus, the  action is written in terms of curvature invariants~\cite{Dubovsky:2012sh, Aharony:2013ipa}, 
\beq
\!A=\int  d^2\sigma \sqrt{-h} \left[\ell_s^{-2} +  \mathcal{R} + K^2 + \ell_s^2K^4+  \dots \right] \, ,  \label{lag}
\eeq
where $\mathcal{R}(h_{\alpha\beta})$ is the Ricci scalar and $K^\mu_{\alpha\beta}=\nabla_\alpha\partial_\beta X^\mu$ is the extrinsic curvature tensor and implicit are Wilson coefficients multiplying any of these structures in the effective Lagrangian.
The parameter $\ell_s$ is  called the string length.
In static gauge $X^\mu(\sigma)=(\sigma^\alpha , \,  X^i)$, where~$i\,{=}\,1, {\dots}, D{-}2$ are the
transverse excitations of the flux tube.

Nicely, Ricci is a total derivative and $K^2$ vanishes on-shell so the first two terms in the effective field theory expansion can be dropped. 
 Therefore, the low energy dynamics of  \rf{lag} is tightly constrained by the non-linearly realized target Poincar\' e symmetry. This is known as  low energy universality~\cite{Dubovsky:2012sh,Aharony:2013ipa,Luscher:2004ib}. 
 The leading deviations from the Nambu-Goto predictions for physical observables arise from $K^4$ operators in~\rf{lag}, namely effects of~$O(\partial^8X^4)$. More precisely, there are two $K^4$ operators, differing  by the contractions of the indices and correspondingly two coefficients $\alpha_3$ and $\beta_3$ which do depend on the specific underlying confining theory. 
 
 We will constrain them in this paper and thus bound interesting physical quantities which depend on them. 
To constrain these parameters we  turn to the  on-shell approach to the flux tube world-sheet theory pioneered by \cite{Dubovsky:2012sh} which is based on the branon \textit{S-matrix}. 
  The~$2\to 2$ scattering amplitudes can be decomposed into channels, i.e.  irreducible representations of the  symmetry group~$O(D{-}2)$. 
The low energy expansion of the phase shifts in each channel can be written as  
(see \ref{Ssetup}  for details)
\begin{align} 
	2\delta_{sym} &=  \frac{s}{4}   + \alpha_2  s^2 + 
	\alpha_3 s^3+O(s^4) \nonumber \\
	2\delta_{anti}  &=  \frac{s}{4}- \alpha_2 s^2 + (\alpha_3 {+}2\beta_3) s^3 +O(s^4) \label{lowenergyexpansion}\\
	2\delta_{sing} &= \frac{s}{4}- (D{-}3)\alpha_2 s^2 +(\alpha_3 {-}(D{-}2)\beta_3) s^3 +O(s^4) \nonumber
\end{align}  
where $\alpha_2=\frac{D-26}{384\pi}$ and $s$ is the square of the center of mass energy.
Here and  below we set $\ell_s=1$.
The low energy universality mentioned above is manifest here up to~$O(s^2)$ included~\cite{Dubovsky:2012sh}. 
The non-universal~$K^4$ terms in~\eqref{lag} contribute at~$O(s^3)$ and are encoded in the parameters~$\alpha_3$ and~$\beta_3$ in~\eqref{lowenergyexpansion}.
The high degree of universality of the branon S-matrix is to be contrasted with a theory of  compact goldstones like the Pion, whose $S$-matrix shows departures from universality  already at $O(s^2)$. To this order, the phase shifts are real because inelastic processes like $2\to  4$ give rise to imaginary contributions of $O(s^6)$.

We will see below that by requiring a consistent UV completion of the branon S-matrix, we can put bounds on its low energy expansion and thus bound the effective field theory parameters. This immediately leads to many interesting bounds on various low energy physical observables. One such interesting observable is the finite volume energy spectrum which we can compute in perturbation theory from the action~\eqref{lag} above. For example, for the ground state, we will find 
 \beq
E_0(R) = \sqrt{ R^2-\tfrac{\pi}{3}(D-2)} + \frac{\delta(D)}{R^7}  + O(1/R^9 ) \, , \label{spec}
\eeq
 where $R$ is the length of the  flux tube loop and
 \beq
\delta(D)= \frac{32\pi^6(2-D)((D{-}2)\alpha_3{+}(D{-}4)\beta_3)}{225} \,.\la{deltaD}
 \eeq
Note that the leading confining potential $E_0(R)\sim R$ all the way upto the sub-sub-sub-leading corrections of~$O(1/R^5)$ are universal and captured by the square root term in (\ref{spec}).

Similar formulae governing the first few universal terms of the large $R$ expansion can be written for excited states as well \cite{Dubovsky:2012sh,Aharony:2013ipa}. 
A particular feature of those results is that they exhibit quite a lot of degeneracy: for very large radius the energy levels typically depend only on the total left and right moving momentum but not on the individual momenta of the branons.
Level splitting of these energy levels starts at $O(1/R^7)$ and directly probes the non-universal parameters $\alpha_3$ and $\beta_3$ introduced above (see equation  \eqref{splitting42} below for a concrete example).

In summary, at very low energy, i.e. very long flux tubes, universality powerfully constrains everything.  Eventually non-universal terms kick in. We then have a triangle of three important players: the effective field theory Lagrangian~\eqref{lag}, the branon S-matrix~\eqref{lowenergyexpansion} and the finite volume spectrum~\eqref{spec}.
The main result of this paper is a bound on the low energy expansion of the S-matrix following from the existence of its consistent UV completion. This immediately translates into rigorous bounds for the other two players in the triangle. 

More speculatively, we will also study the boundary of the allowed S-matrix space and find a remarkable numerical coincidence: on that boundary lies an S-matrix exhibiting a resonance with the quantum numbers, mass and width exactly as predicted in~\cite{Dubovsky:2013gi,Dubovsky:2014fma} and dubbed as the QCD worldsheet axion there. Amusingly, at that same point, the S-matrix we obtain also contains three further heavier excitations which we call the dilaton, the symmetron and the axion*. Given the remarkable numerical coincidence w.r.t. the QCD axion, it is tempting to speculate that they should be present in the QCD flux tube. 

Lattice Monte Carlo simulations of pure Yang-Mills provide precious information on the dynamics of confining flux tubes. The measurements of the low energy spectrum support the outlined picture of universality at large radius --  see~\cite{Teper:2009uf} for a review -- and should hopefully be sensitive to the non-universal corrections soon, e.g.~\cite{Athenodorou:2016kpd} for~$D{=}3$. They also favor the existence of the conjectured axion excitation~\cite{Athenodorou:2010cs,Dubovsky:2013gi,Dubovsky:2014fma,Athenodorou:2017cmw}; it would be very interesting to look for other more massive excitations.

\section{2D massless S-matrix Bootstrap}
\label{sec2}
Massless excitations in 2D can be left (L) or right (R) movers.
In this section, we study the L-R scattering amplitude of branons.\footnote{A general 2D massless S-matrix has non-trivial  L-L, L-R and R-R components~\cite{Fendley:1993xa}.  
Lorentz invariance implies that the L-L components is a function of the ratio $p^L_1/p^L_2$ of the momenta of the incoming left-moving particles. 
This means that the L-L and the R-R amplitudes are independent of the energy scale.
Therefore, the branons must have trivial L-L and R-R scattering because their interactions turn off at low energies~\cite{Dubovsky:2012wk}.}

\subsection{Setup} \label{Ssetup}
A long flux tube in $D$ dimensions breaks the target Poincar\'e symmetry to $ISO(1,1)\times O(D-2)$.\footnote{We assume that the $D$-dimensional theory and the flux tube preserve parity. It would be interesting to generalize our study of branon scattering in the absence  of parity, e.g. due to a $\theta$-term. } This leads to $D-2$ Goldstone bosons or branons.
Consider now the $2\to 2$ scattering amplitude of these branons,
\beq \label{4dsmatrixdecomp}
S_{ab}^{cd}(s)=\sigma_1(s) \delta_{ab}\delta^{cd}+\sigma_2(s)\delta_a^c\delta_b^d+\sigma_3(s)\delta_a^d\delta_b^c\, ,
\eeq
where the indices run over the $D-2$ transverse directions and  $s$ is the square of the center-of-mass energy. Crossing symmetry leads to
\begin{eqnarray}\label{4dcrossing}
\sigma_2(-s) = \sigma_2(s) \,,\qquad \qquad
\sigma_3(-s) = \sigma_1(s) \,.
\end{eqnarray}
The amplitude~\eqref{4dsmatrixdecomp} can also be decomposed in partial waves of $O(D{-}2)$ namely, the singlet, the anti-symmetric tensor and the symmetric traceless tensor 
(see \cite{Cordova:2018uop,He:2018uxa} for details), 
\begin{eqnarray}\label{isospinamplitudes}
S_{sing} &=& e^{2i \delta_{sing}}=(D-2) \sigma_1 + \sigma_ 2 + \sigma_3 \nonumber \\
S_{anti} &=& e^{2i \delta_{anti}}=\sigma_2 - \sigma_3 \\
S_{sym} &=& e^{2i \delta_{sym}}=\sigma_2 + \sigma_3 \nonumber 
\end{eqnarray}
where $\delta_{rep}$ 
may have an imaginary part due to particle production.
In this basis, unitarity is simply 
\beq
|S_{rep}(s)|^2 \leq 1\, ,  \quad \forall  s > 0 \, .
\eeq
The amplitudes $\sigma_i(s)$ are analytic functions of $s$ in the upper and the lower half plane related by~\footnote{This is just real analyticity for massive particles. For massless particles, the $s$-channel and the $t$-channel cuts touch at $s=0$ and cover the entire real axis of $s$.}
\beq \label{realanalyticity}
\sigma_i(s^*) = \[\sigma_i(s)\]^*\,.
\eeq
Therefore, it is enough to know the amplitudes in the upper half plane, where
equations~\eqref{4dcrossing} and~\eqref{realanalyticity} lead to
\begin{eqnarray}\label{smatrixsplane}
\sigma_2(-s^*)=\left[\sigma_2(s)\right]^*\,,\qquad  
\sigma_3(-s^*)=\left[\sigma_1(s)\right]^*\,.
\end{eqnarray}

The Nambu-Goto lagrangian leads to the low energy expansion of the phase shifts  as $2\delta_{rep} =\frac{s}{4}+O(s^2)$.
In principle, higher order terms may also include non-analytic  terms of the form $s^p (\log s)^k$ with $p>k> 0$.
Furthermore, we know that $\text{Im}\,\delta_{rep} = O(s^6)$ because particle production starts with $|\mathcal{M}_{2\to 4}|^2 \sim  l_s^{12} $ \cite{Cooper:2014noa}.
Using just these facts and \eqref{smatrixsplane} we can derive the low energy expansion \eqref{lowenergyexpansion} with $\alpha_2$, $\alpha_3$ and $\beta_3$ as real parameters.
In the context of the flux tube theory $\alpha_2=\frac{D-26}{384\pi}$ is universal and $\alpha_3$ and $\beta_3$ are non-universal coefficients related to the two independent  $K^4$ terms in \eqref{lag}.
In appendix~\ref{ap:deltaexpansions}, we push this expansion up to $O(s^6)$ and find perfect
agreement with the $O(s^4)$ results of \cite{Conkey:2016qju}.

\subsection{$D=3$ Flux Tubes}
\label{ftd3}

\begin{figure}[t]
\centering
        \includegraphics[scale=0.34]{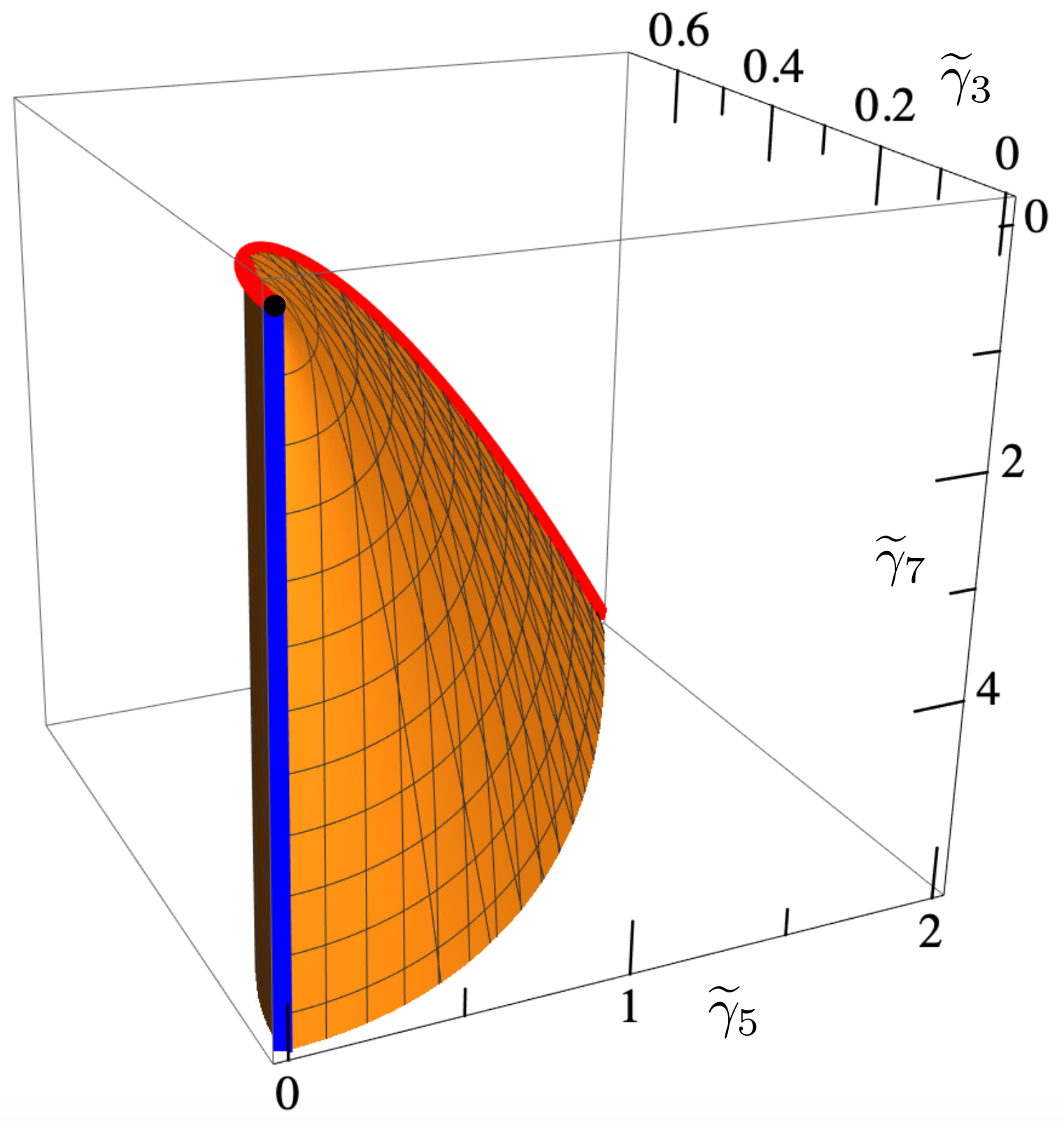}
 \caption{ Allowed region in the $\{\tilde\gamma_3,\tilde\gamma_5,\tilde\gamma_7\}$ space for a generic $D{=}3$ flux tube S-matrix, with $\tilde\gamma_n{=}\gamma_n+(-1)^{(n+1)/2}\frac{1}{n 2^{3n-1}} $.
 The S-matrix at the cusp (black point) is associated to the goldstone (goldstino) S-matrix describing the flow from tricritical Ising to free fermions: it saturates the Schwarz-Pick inequality. The edge in red corresponds to double CDD solutions, saturating the 2-point Schwarz-Pick bound and the full orange surface is determined by the 3-point Schwarz-Pick inequality and it is saturated by a triple CDD family.} 
\label{3Dfigure}
\end{figure}

To start with, we focus on the $D=3$ target space. 
In this case, only $\delta_{sing}\equiv \delta$ is meaningful and the amplitude~$S=e^{2i\delta}$
obeys $S(-s^*) = \[S(s)\]^*$ for $s$ in the upper half plane. Furthermore, it was shown in~\cite{Chen:2018keo} that~$\Im \delta =O(s^8)$. This implies
\beq \label{lee3D}
2\delta(s) = \frac{s}{4}   + \gamma_3 s^3 + 
\gamma_5 s^5+\gamma_7 s^7+i\gamma_8 s^8 +O(s^9)  \, , 
\eeq
where $\gamma_3, \gamma_5, \gamma_7$ are non-universal parameters.
On the other hand, $\gamma_8$ is determined by  the probability of particle production 
at leading order
$
P_{2\to n\ge 4} = 2\gamma_8 s^8 + O(s^9)
$
. 
As explained in \cite{Chen:2018keo}, $\gamma_8 \propto \gamma_3^2$ is not an independent parameter. We shall now show that the coefficients~$\gamma_3, \gamma_5, \gamma_7$ can only take values in the region depicted in figure \ref{3Dfigure}.

The S-matrix $S(z)$ is a holomorphic function from the upper half plane $\mathbb H$ to the the unit disc $\mathbb D$ because unitarity on the real axis along with the maximum modulus principle implies that $|S(z)|
\le1$ in the full upper half plane. Next, we construct a new function
\beq
S^{(1)}(z|w) \equiv \frac{S(z)- S(w) }{1- S(z) \overline{S(w)}} \Big/\frac{z-w}{z - \overline w} \, ,    \label{SP1}
\eeq
where $w$ is any point in the upper half plane.
It is easy to see that (as a holomorphic function of $z$) this function $(a)$~has no singularities in the upper half plane and $(b)$~is again bounded by 1 for $z$ on the real line.\footnote{For $z$ on the real line, $|z-w|/|z-\overline w| = 1$ and notice that $S(z)$, $S(w) \in \mathbb D$ and $\frac{S(z)-S(w)}{1-S(z)\overline{S(w)}}$ is a Mobius transformation that maps the unit disc $\mathbb D$ to itself.} By the maximum modulus principle, it is bounded everywhere on the upper half plane: $|S^{(1)}(z|w)|_{\Im(z)\,\ge\, 0} \le 1$. This is the content of the so-called Schwarz-Pick theorem.

Inserting \eqref{lee3D} in the Schwarz-Pick combination \eqref{SP1}
and expanding  for small and imaginary $z$ and $w$, we find
\beq
S^{(1)}(i x |i y)=-1+\left(\frac{1}{96}+8 \gamma_3\right) x\,y+\dots \geq -1 \, .
\eeq
This leads to our first  bound  
\beq
\gamma_3 \geq -\frac{1}{768}\,.
\eeq
In Appendix~\ref{appendix:SP}, we show that the bound cannot be improved by approaching the origin $z{=}w{=}0$ in any other direction in the upper half plane.
The authors of~\cite{Dubovsky:2014fma, Chen:2018keo} estimated~$\gamma_3 \approx 3\times 10^{-4}$ from lattice data for $SU(6)$ YM~\cite{Athenodorou:2011rx}.  

\begin{figure}[t]
\centering
        \includegraphics[scale=0.3]{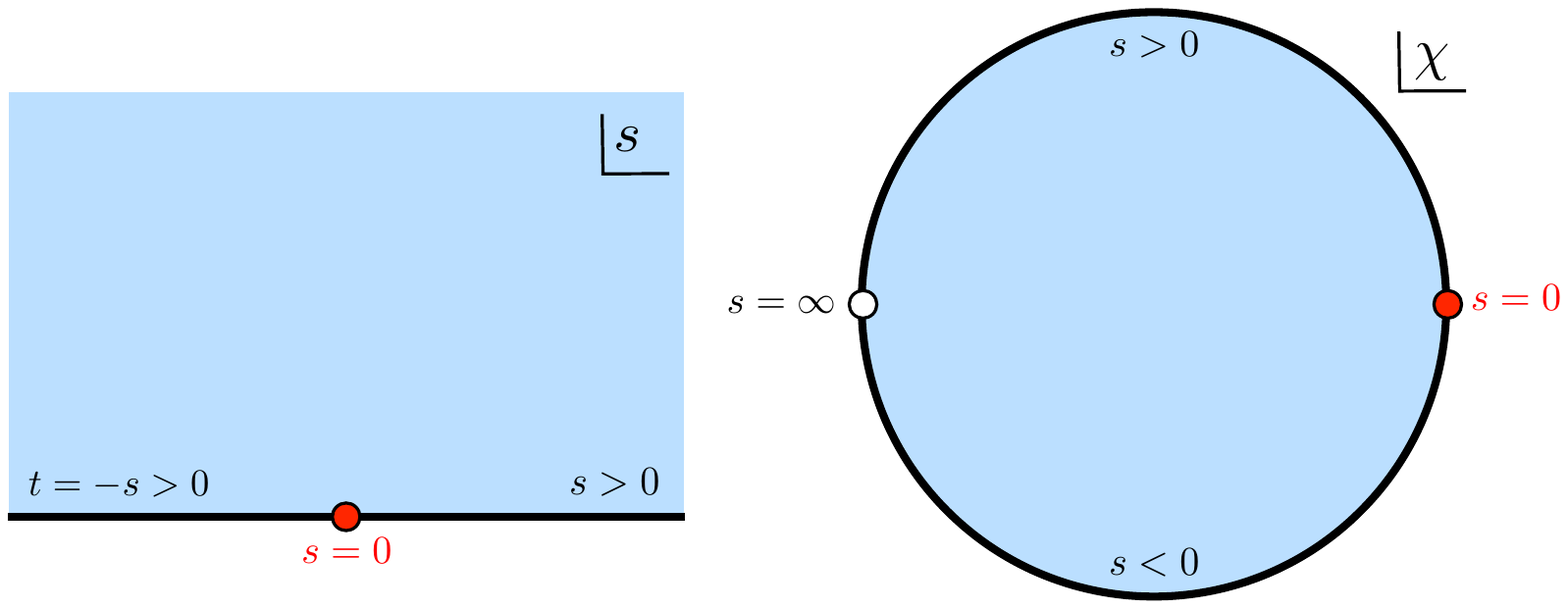}
\vspace{-3cm} \caption{Left: domain of analyticity of a generic massless two-dimensional S-matrix. The cut, in black, is all over the real axis; the threshold at $s=0$ it is in general a branch point singularity. Right: we map the upper half plane to the unit disc through $s\to \chi=(4+is)/(4-is)$. The real axis is mapped to the boundary of the unit circle, the threshold to $\chi=1$ and~$s=\infty$~to $\chi=-1$. } 
\label{unitDisk}
\end{figure}

Similarly, one can define $S^{(2)}(z|q,w)$ by replacing $S(z)$ by $S^{(1)}(z|q)$ in~\eqref{SP1}.
Such Schwarz-Pick multi-point generalizations~\cite{BeMi}
can be used to derive (see appendix~\ref{appendix:SP})
\beqa
\tilde  \gamma_3&\geq& 0\nonumber \\ 
\tilde \gamma_5&\geq& 4 \tilde \gamma_3^2 - \frac{1}{64}\tilde \gamma_3 \label{3DSP}\\
\tilde \gamma_7&\geq &\frac{\tilde \gamma_5^2}{\tilde \gamma_3}+\frac{1}{4096}\tilde \gamma_3+\frac{1}{64}\tilde \gamma_5-\frac{1}{16}\tilde \gamma_3^2
\nonumber
\label{rockinequalities}
\eeqa
where $\tilde\gamma_n{=}\gamma_n+(-1)^{(n+1)/2}\frac{1}{n 2^{3n-1}}$.
The allowed region is shown in figure~\ref{3Dfigure}. 

It is interesting that Schwarz-Pick inequalities exploit both unitarity and analyticity by exploring the region of purely imaginary Mandelstam $s$ -- orthogonal to real physical $s$ -- to efficiently bound the space of $2\to 2$ S-matrices.
 
\subsection{$D=4$ Flux Tubes}  \la{D4section}

In $D=4$ dimensions, the branon S-matrix possesses an $O(2)$ symmetry. In addition, the crossing and unitary equations are invariant under  $S_{\text{sing}} \leftrightarrow S_{\text{anti}}$ interchange corresponding to $\beta_3 \leftrightarrow {-}\beta_3$ in  \eqref{lowenergyexpansion}. Universality fixes the low energy expansion of the phase shifts up to order $s^2$ included.
The leading non-universal behavior depends on the two coefficients $\alpha_3$ and $\beta_3$ introduced in \eqref{lowenergyexpansion}.

Crossing mixes the various irreps but the symmetric channel S-matrix is still bounded by $1$ along all the real $s$-axis\footnote{By crossing $|S^{\text{crossed}}_{\text{sym}}|=\frac{1}{2}|S_{\text{sing}}+S_{\text{anti}}|\leq 1$. This actually holds for any $O(N=D-2)$ theory as pointed out  in \cite{Cordova:2018uop}.}
so we can still apply the first Schwarz-Pick inequality in this channel as in the previous section. 
  Moreover, it can be applied to the two crossing symmetric combinations in $D=4$: $S_\pm=1/2(S_{sing}\pm S_{sym})$.
This analysis leads to 
\beqa
\alpha_3 \geq -\frac{1}{768}+\frac{121}{9216 \pi^2}\, ,\\
\alpha_3 \geq -\frac{1}{768}+|\beta_3|\,.
\label{SPSym}
\eeqa

This is however not the full allowed $\{\alpha_3,\beta_3\}$ space as we have yet to explore all channels and their interrelations. To find the optimal bounds we proceed numerically in the spirit of~\cite{Paulos:2016but, Paulos:2017fhb, Guerrieri:2018uew}.
We map the upper half plane to the unit disk with the real axis mapped to the unit circle, see figure \ref{unitDisk}.
By assumption the S-matrix is analytic in the interior of the disk and we can represent it as a Taylor expansion.
Our numerical ansatz is then given by the truncated Taylor series 
\beq
S_{\text{ansatz}}=\sum_{n=0}^{N_{max}} a_n \chi^n\, ,\quad \quad |\chi|\leq 1 \, .
\eeq
Then we minimize the linear functional $\alpha_3$ in the vector space of the Taylor coefficients $\{a_n\}$ as a function of~$\beta_3$, given the quadratic constraints $|S_{\text{ansatz}}(\chi)|\leq 1$ for each $\chi$ on the upper boundary of the disk and for each~$S_{rep}$.
Further details are given in appendix \ref{ap:numerics} along with more general numerical results obtained as byproduct of our explorations but that are not relevant in the context of flux tube theories.

\begin{figure}[t]
\centering
        \includegraphics[scale=0.24]{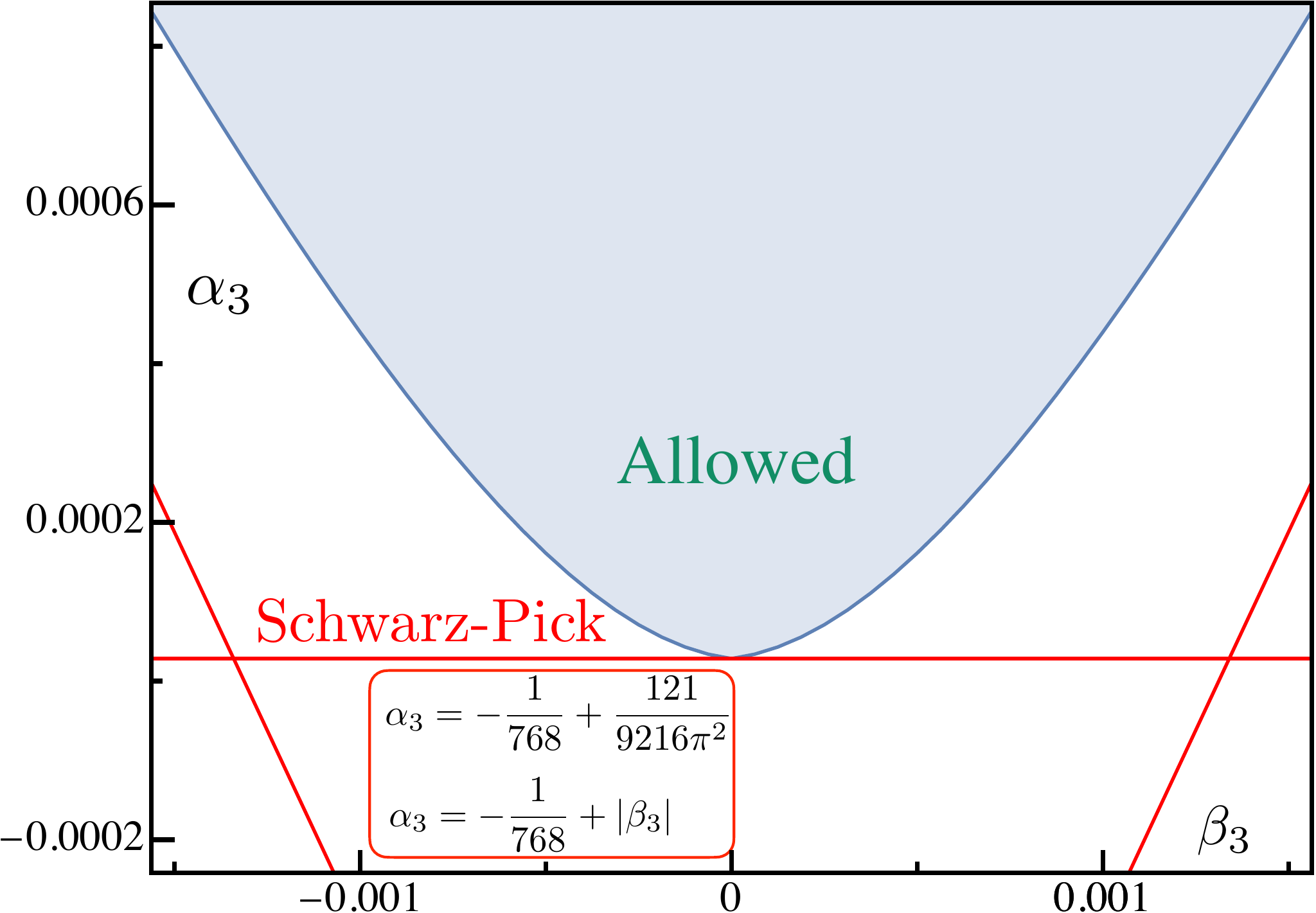}
 \caption{ Allowed region in the $\{\beta_3,\alpha_3\}$ parameter space of flux tube S-matrice in $D=4$ as obtained by numerics. 
 The horizontal red line represents the absolute minimum of $\alpha_3$ as predicted analytically by the Schwarz-Pick theorem applied to the symmetric channel.
  The additional red lines can be obtained applying Schwarz-Pick to the additional crossing symmetric combinations $S_\pm(s)$.
 }
\label{fig4dboundary}
\end{figure}

The numerical result of the optimization problem is shown in figure~\ref{fig4dboundary}.
The analytic bound in~\eqref{SPSym} would allow all the points above the Schwarz-Pick line (in red), while we see numerically that the effect of bounding the other 
channels produces the region depicted in blue.
When~$\beta_3=0$ the numerical bound and the analytic one coincide. At this point,
the S-matrix satisfies Yang-Baxter and it is a pure phase in all channels.
Its expression can be predicted analytically and is given in appendix \ref{YBAppendix}.

\section{Energy spectrum in finite volume }
\label{ens}

We just saw how many constraints on the long Flux Tube follow from the general principles of the S-matrix theory.  Here we translate them into  constraints on large volume    observables. We start with the flux tube ground state energy $E_0(R)$. 

At very large $R$ we read off the string tension from~$E_0(R) \simeq R/\ell_s^2$. Recall that the  corrections up to~$1/R^5$ to this confining result are  universal  and given by the square root   in   (\ref{spec}).  The sub-sub-sub-subleading term is \textit{not} uniquely fixed by symmetry and is the subject of this section.

Computing the non-universal correction in \rf{spec} is straightforward in perturbation theory (sum of connected vacuum Feynman diagrams),  albeit increasingly complex as we move to higher orders in $1/R$. The leading correction $\delta(D)$  comes from the two $K^4$ possible interactions which we parametrize as 
\beqa \label{lagalphbet}
&&\mathcal{L}_\text{non-univ}= \partial_a \partial_b X^{i} \partial_a \partial_b X^{j} \partial_c \partial_d X^{k} \partial_c \partial_d X^{l} \times \\&&\qquad\qquad\qquad \times  \[ 4\delta_{ik} \delta_{jl} (\alpha_3+ \beta_3)-2\delta_{ij} \delta_{kl}  (\alpha_3+3 \beta_3)\]\nn \,. 
\eeqa
Here we parametrize the coefficient of the two invariant structures so to match the eye pleasing expressions (\ref{lowenergyexpansion}), as can be verified by a straightforward tree level computation.
Thus, the leading order non-universal contribution to the vacuum energy density is 
  \beq
\begin{minipage}[h]{0.12\linewidth}
        \vspace{0pt}
        \includegraphics[width=\linewidth]{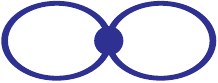}  
   \end{minipage}   =f(D) \left[\partial_\mu  \partial_\nu \partial_\rho\partial_\sigma \Delta_R(0) \right]^2 \, ,
   \label{devb1}
      \eeq
      where $f(D)
     = 4(2-D)((D{-}2)\alpha_3{+}(D{-}4)\beta_3)$.
     The derivative of the finite volume propagator is given by $\partial_\mu \Delta_R (x){=}  \sum_n \partial_\mu \Delta(x+n)$, where~$\partial_\mu \Delta(x) = -x_\mu/(2\pi \,  x^2)$ and $n_\mu =(0,n R)$ is a displacement vector in the winding direction. 
     The zero mode $n_\mu =(0,0)$  gives a short-distance divergence in the limit $x \rightarrow 0$ leading to  \rf{devb1}. This is regulated by a local counter-term, which at this order simply amounts to neglecting the zero mode. Thus, after a bit of algebra and excluding the zero mode,  we are led to 
     $ \left[\partial_\mu  \partial_\nu \partial_\rho\partial_\sigma \Delta_R(0) \right]_\text{ren.}^2  = 288/\pi^2\sum_{n,m=1}^\infty 1/(R^8n^4 m^4){=}8 \pi^6/225R^8 $, which gives the desired relation \rf{deltaD} between the first non-universal correction to the energy and the first non-universal low energy S-matrix parameters or Wilson coefficients. See appendix \ref{perturbativeAppendix} for further details. 
(In~$D=3$, physical quantities  only depend on the combination  $\alpha_3-\beta_3=\gamma_3$.)
     
Since we bounded the later low energy parameters, see figures \ref{3Dfigure} and \ref{fig4dboundary}, we automatically obtain bounds on the Wilson coefficients and on the ground state energy. In three and four dimensions, for instance, we find the following bound on the deviation from the square root formula
\beq
\delta(3) =- \frac{32\pi^6 \gamma_3}{225} \leq \frac{\pi^6}{5400} \, , \la{nonunivbounds3D}
\eeq
and
\beq
\delta(4) =-\frac{128\pi ^6 \alpha_3}{225}  \leq \frac{\pi^6}{1350} - \frac{121\pi^4}{16200}  \, .  \la{nonunivbounds4D}
\eeq
Note that the right hand side of (\ref{nonunivbounds4D}) is negative so the square root formula \textit{must} be corrected; the right hand side of (\ref{nonunivbounds3D}) is positive, 
in nice agreement with the 
fact that integrable $D = 3$ Strings have precisely
 $E_0^\text{int}= \sqrt{ R^2-\tfrac{\pi}{3}}$.
Note also that the four dimensional bound (\ref{nonunivbounds4D}) is saturated when $\beta_3=0$ (see figure \ref{fig4dboundary}) which corresponds to the particular point where integrability is preserved.

 In fact, if we exploit the low energy integrability of the theory we can bypass the Lagrangian approach altogether and, by means of the so called Thermodynamic Bethe Ansatz (TBA), compute \rf{spec} in terms of the S-matrix. 
 
This is particularly clean in $D=3$ since there is only a single branon and a single corresponding pseudo-energy in this case. The ground state energy then reads 
\beq
{E}_0(R)=R+\frac{1}{\pi R}\int_0^\infty dx\,\log(1-e^{-\varepsilon(x)})  \, , 
\label{Etba}
\eeq
where the pseudo-energy $\varepsilon$ is the solution to the integral equation
\beq
\varepsilon(x)=x+\frac{1}{2\pi} \int_0^\infty \frac{dx^\prime}{x^\prime}\mathcal{K}
\log(1-e^{-\varepsilon(x^\prime)}) \, , 
\label{tba}
\eeq
with the kernel $\mathcal{K}=x^\prime\! \tfrac{\partial}{\partial x^\prime}\,\delta(\tfrac{4 x x^\prime}{R^2})$ and the phase shift is given by the low energy expansion  (\ref{lee3D}). (The TBA aficionado might notice that this equation is a bit unusual; in terms of $x=e^{\theta}$ we see that a \textit{sum} of rapidities shows up as opposed to the more conventional \textit{difference}. This is because of the pure $L/R$ scattering of our problem.)  Expanding the pseudo-energy as 
\beq
\varepsilon(x)=x+\frac{a(x)}{R^2}+\frac{b(x)}{R^4}+\frac{c(x)}{R^6}+\mathcal{O}\left(\frac{1}{R^8}\right) \, , 
\label{pseudoansatz}
\eeq
and collecting powers of $R$ we 
can straightforwardly find all the functions $a,b,c,\dots$ and hence the ground state energy,
\beq
E_0= R{-}\frac{\pi}{6 R}{-}\frac{\pi^2}{72 R^3}{-}\frac{\pi^3}{432 R^5}{-}\frac{5\pi^4}{10368 R^7}-\frac{32\pi^6 \gamma_3}{225 R^7}{+}\dots  \la{finalE} 
\eeq
We recognize that the first five terms are precisely the expansion of the square root in~\eqref{spec} while the last term is nothing but~\eqref{deltaD} with $D=  3$ and $\alpha_3-\beta_3\equiv \gamma_3$.  

For $D=4$ we can proceed in the same fashion as long as we restrict ourselves to the integrable subspace $\beta_3=0$. In this case, we get a  set of TBA equations~\cite{Dubovsky:2014fma} which can also be  simplified into a single equation and solved in a large $R$ expansion, see appendix~\ref{appendixtba} for details.

Next we have excited states which we can analyze in a similar way, either through perturbation theory or through the \textit{excited state} TBA. Of particular relevance is the level splitting between the first few energy levels since this degeneracy lifting is a sharp signature of the non-universal terms. 
For instance, in $D=3$  the first level splitting between the second and third excited states (in the zero momentum sector) reads 
 \beqa
&&{E}_\text{2 branons}(2,-2)-{E}_\text{4 branons}(1,1,-1,-1) \nn \\&&\qquad =-\frac{2455552\pi^6 \gamma_3}{5 R^7} \,+\, O \left({1}/{R^9}\right) \la{splitting42}\, ,
\eeqa
where the arguments refer to the individual mode numbers.
The basic logic that goes into computing \rf{splitting42} is analogous to the derivation of~\rf{finalE}, therefore the details are given in appendix~\ref{LevelSplitting}. There we also establish the bound on~\eqref{splitting42}, that immediately follows from~\eqref{3DSP}, and comment on potential comparisons with LMC data in the future. 

Note that in our current logic, we can not completely ignore the Lagrangian since we are exploiting the thermodynamic Bethe ansatz, which only allows us to relate the S-matrix and the energy levels when the system is integrable. It should be possible -- and very interesting -- to relate more generally the various energy levels with the two-to-two S-matrix, together with all higher point amplitudes of non-integrable theories. The L\"{u}scher corrections~\cite{Luscher:1990ux} provide the leading term and generalized L\"{u}scher corrections have been recently explored e.g. in~\cite{Briceno:2017tce}.  Perhaps the recent rederivation of the TBA in more diagrammatic terms, see e.g.~\cite{Kostov:2018ckg}, can provide some insights for such putative description. Or,  developing the approach of~\cite{Dashen:1969ep} for the flux tube may turn out useful.
It would be great to adapt these ideas to our setup and re-derive~\eqref{nonunivbounds4D} without expanding around the integrable theory.

\section{Resonances}
\label{section:resonances}
Given the bounds in figures~\ref{3Dfigure} and~\ref{fig4dboundary} it is natural to ask which S-matrices lie on those boundaries. This is particularly relevant in 4D since Lattice MC shows a  rich phenomenology, with the presence of a  parity odd resonance~\cite{Athenodorou:2010cs,Athenodorou:2017cmw},  dubbed  QCD worldsheet axion in the S-matrix approach to the long flux tube~\cite{Dubovsky:2014fma,Dubovsky:2013gi}.

In $D=3$ we can find the S-matrices at the boundary of    figure~\ref{3Dfigure} analytically. Given that in this case there is no strong evidence for resonances from the lattice data, we present this analysis in appendix \ref{SPapplication}. 
Furthermore, ref.~\cite{Dubovsky:2015zey}  suggests  that indeed  the $D=3$ QCD Flux-Tube has no resonances,  in appendix~\ref{asa} we discuss how our bounds can be improved if we incorporate this further assumption into the analytic properties of the  S-matrix. 
Finally, let us  mention that at the cusp  of  figure~\ref{3Dfigure} $S_\text{cusp} = (8i-s)/(8i+s)$. Nicely, this is an important S-matrix, albeit in a different context: it describes the RG flow from the tricritical Ising fixed point to the free fermion theory~\cite{Zamolodchikov:1991vx}.~\footnote{Precisely, 
$S_\text{Goldstino}=-S_\text{cusp} $, where the  overall minus  sign could be easily incorporated in the formulas of the main text.}

\begin{figure}[t]
\centering
        \includegraphics[scale=0.276]{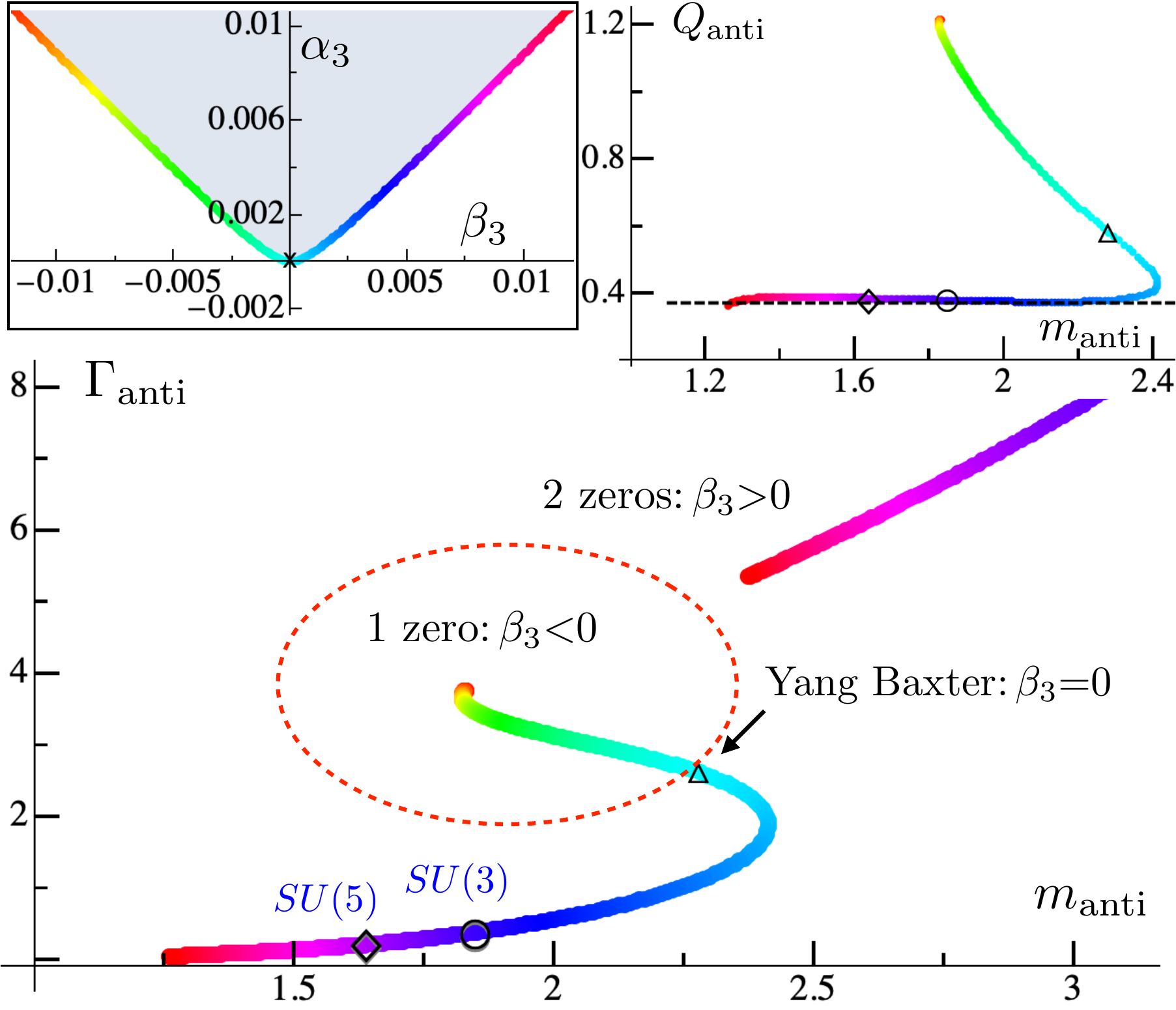}
 \caption{At the S-matrix space boundary we encounter S-matrices with zeros, that is resonances. In the antisymmetric channel, to the left of the integrable point there is one single resonance while to the right of the integrable point there are two resonances, a broad one and a sharp one. Curiously, as we move along the boundary we encounter S-matrices whose resonance mass and with are in precise agreement with those predicted in~\cite{Dubovsky:2014fma,Dubovsky:2015zey} as extracted from $SU(3)$ (right point) and $SU(5)$ (left point) lattice data.}
\label{figaxion}
\end{figure}

Next we turn to $D=4$, where the boundary must be studied numerically. Here we find some remarkable surprises. A first nice surprise is that the S-matrices which saturate the bound have zeros,  which physically correspond to resonances. Figure \ref{figaxion} describes the position of these resonances in the anti-symmetric channel as we move along the boundary of figure \ref{fig4dboundary}. Depending on whether we are to the right or left of the integrable $\beta_3=0$ point, there is one ($\beta_3<0$) or two zeros there  ($\beta_3>0$). As we move along the boundary in the region  $\beta_3>0$, the sharpest of these resonances passes spot on by the values of the worldsheet axion. The two dots correspond to estimates based on $SU(3)$ \cite{Dubovsky:2014fma,Dubovsky:2013gi}  and $SU(5)$  \cite{Dubovsky:2015zey} lattice MC simulations \cite{Athenodorou:2010cs}. 
Because of these encouraging numerical coincidences we will   denote these two points along the boundary as the $SU(3)$ point and $SU(5)$ point.

Remarkably, at these points, we find other zeros in the S-matrices. One broader resonance shows up in the same anti-symmetric channel along with a resonance in each of the other two channels. 
The spectrum, measured as $s_0=(m+i\Gamma/2)^2$ at the position $S_{rep}(s_0)=0$, 
for the~$SU(3)$ and~$SU(5)$ points is given by
\begin{center}
\begin{tabular}{L{3.1cm}L{2.4cm}L{1.67cm} }
\specialrule{.7pt}{1pt}{1pt}
  \textbf{spectrum $\bm{ [m,\Gamma]}$} & $\bm{SU(3)}$ & $\bm{SU(5)}$  \\[.05cm]
\specialrule{.4pt}{1pt}{1pt}
     axion &  $[1.85,\, 0.39]$ &    $[1.64,\,0.22] $ \\
   axion* &  $[3.25,\,8.84]$   &  $[2.83 ,\,7.02]$ \\
   symmetron &  $[2.36,\,4.99]$   &  $ [2.34,\, 4.54]$ \\
   dilaton  & $[1.88,\,3.37]$  & $[1.84,\, 3.52]$ \\
\specialrule{.7pt}{1pt}{1pt}
\end{tabular}
\label{spectrum}
\end{center}
Even though these values should obviously be taken as benchmark values only,  
could these resonances be further  excitations present in the  Yang-Mills long flux tubes?
Of course, these explorations must be taken with a grain of salt since there is a priori no strong reason for the real flux tube to be close to the boundary (recall that  these S-matrices cannot represent the elastic scattering of branons all the way up to the UV since particle production kicks in eventually~\cite{Cooper:2014noa}). 

Furthermore,  we find that the axion coupling to the world-sheet branons $Q\approx\sqrt{8\Gamma}/m^{5/2}$ (valid for $\Gamma \ll m$) 
shows a plateau for almost all the values of $\beta_3>0$, apart from an initial transient. The value of this plateau surprisingly coincides within a good numerical accuracy with~$Q_{\text{integrable}}= \sqrt{7/(16 \pi)}$~\cite{Dubovsky:2015zey}.
The integrable value of~$Q$ is fixed by demanding that the presence of the axion results in a vanishing $2\rightarrow 4$ scattering amplitude as the axion mass goes to zero. 
It is plausible, given our current numerical explorations, that this plateau holds all the way up to $\beta_3\to\infty$, where the axion becomes massless. 
Since the family of S-matrices we find, coincidentally, contain the $SU(3)$ and $SU(5)$ axions, it would be tempting to   believe that all $SU(N)$ axions lie somewhere along the trajectory in figure~\ref{figaxion}
and a natural candidate for the large-$N$ axion would be the massless resonance in the limit $\beta_3\to\infty$. However, this is excluded by current lattice simulations~\cite{Athenodorou:2017cmw} and it 
is an open question how to relate a ``blind to color" S-matrix approach to the~$\text{large-$N$}$ limit.
In appendix~\ref{resonancesAppendix} we give further details on the resonances.

Finally let us mention a small curiosity. The world-sheet axion is the sharpest excitation by far. One could imagine a dilation like excitation, governing the flux tube thickness, to be the lightest scalar mode in some flux tube theories. Could it be that perhaps the broad dilaton identified here becomes sharper in some other circumstances? As it turns out, there is a region where the dilaton and axion swap their roles; it is nothing but the reflection symmetry of figure~\ref{fig4dboundary} discussed in section~\ref{D4section}. So it is  natural to expect  flux tube theories -- albeit with very different physics -- living on the left shore. The left and right shores of figure~\ref{fig4dboundary} can therefore also be called the dilaton and axion shores respectively. In recent QCD explorations~\cite{Guerrieri:2018uew} a boundary of the physical S-matrix of pions was analyzed and there also the two shores had somehow mirror physical properties (as far as which channels are attractive/repulsive, which ones contain the $\rho$ particle etc); the two dimensional setup here highlights in a very sharp way some of the features which are quite challenging to probe numerically there. 

\section{Discussion}

The physics of flux tubes is very rich. At low energy we have universality which constrains very powerfully all physical observables. Here we bounded the first non-universal corrections   by means of the S-matrix bootstrap. In short, the existence of a properly UV completed branon S-matrix constrains the possible space of S-matrices and their low energy expansion, and therefore the space of Wilson coefficients in effective field theory language. 
To our knowledge, this is the first time that the low energy Wilson coefficients are bounded optimally   -- see e.g.~\eqref{3DSP}.
Clearly, our bounds also apply to any string-like defect with the same symmetry breaking pattern as flux tubes.
Moreover, the very same logic could be applied more broadly to bound other systems with spontaneous symmetry breaking in two dimensions such as those arising from broken supersymmetry and even in higher dimensions. It would be fascinating to follow our approach to bound the leading irrelevant operators showing up in pion physics  or, more ambitiously, in gauge and gravity theories; and to compare with existing bounds based on the analytic properties of the forward amplitude~\cite{Adams:2006sv,Komargodski:2011vj,Camanho:2014apa}. We look forward to pursuing this further. 

As we crank up the energy our ignorance grows. At intermediate energy we still have important hints from lattice which strongly suggest the existence of a novel pseudo-scalar like excitations in the flux tube. In S-matrix language such new excitations show up as resonances. Here we looked for the S-matrices living at the boundary of the allowed S-matrix space and found that remarkably they do present a rich resonance pattern and -- what is more -- the masses and width of the the pseudo-scalar resonance pass spot on by the more recent estimates of the putative new particle. Motivated by this numerical coincidence, we are encouraged to take seriously the other broader resonances in  these S-matrices. It would be interesting to look for them on the lattice. 

We could also follow the same sort of games as in~\cite{Guerrieri:2018uew} (where the $\rho$ particle was sometimes imposed as an input) and repeat our Wilson coefficient bound analysis and resonance spectroscopy after fixing the position of the world-sheet axion resonance, to the $SU(3)$ value say, see appendix~\ref{axionAppendix}. 
It would be interesting to push this kind of analysis further, with stronger interplay between the lattice and the S-matrix bootstrap. 

As we increase the energy further we reach an even more mysterious territory. At very high energy we know close to nothing. Do we recover integrability at high energies, that is no particle production and thus~$|S(s) | \to 1$ as~$s \to \infty$. Or, instead, are we  dominated by inelastic particle production with $|S(s)|\to 0$ or even some more exotic possibility? We don't know. It would be very interesting  to figure this out and use it as input  in an improved flux tube S-matrix bootstrap. One possible approach to these questions  would be to include multi-particle amplitudes in the S-matrix bootstrap and constrain those as well. 

\begin{acknowledgments}
We thank C. Bercini, L. Cordova, F. Coronado, L. Di Pietro, S. Dubovsky, J. R. Espi\-nosa, D. Gaiotto, V. Goncalves, V. Gorbenko,  A. Homrich, Z. Komargodski, M. Kruczenski, J. Maldacena, R. Matheus, M. Meineri, R. Myers, A. Patella, R. Rattazzi, M. Riembau, G. Villadoro for useful discussions and comments on the draft. 
Research at the Perimeter Institute is supported in part by the Government of Canada through NSERC
and by the Province of Ontario through MRI. 
This work was additionally supported by a grant from the Simons Foundation (JP: \#488649, PV: \#488661) and FAPESP grant 2016/01343-7 and 2017/03303-1. 
JP and AH are supported by the Swiss
National Science Foundation through the project 200021-169132 and through the National
Centre of Competence in Research SwissMAP.
\end{acknowledgments}

\appendix

\section{Low energy expansion}
\label{ap:deltaexpansions}

Loop diagrams in the branon effective field theory  may lead to non-analytic  terms of the form $s^p (\log s)^k$ with~$p>k> 0$.
Thus, we consider the following general low energy expansion
\begin{align}
\sigma_1 &=  \sum_{p=1}^6 \sum_{k=0}^{p-1} (a_{p,k}+i b_{p,k}) (is)^p \[\log (-is)\]^k
\nonumber\\
\sigma_2 &=1+  \sum_{p=1}^6 \sum_{k=0}^{p-1} c_{p,k} (is)^p \[\log (-is)\]^k
\label{leansatz}\\
\sigma_3 &=  \sum_{p=1}^6 \sum_{k=0}^{p-1} (a_{p,k}-i b_{p,k}) (is)^p \[\log (-is)\]^k
\nonumber
\end{align}
where the coefficients $a$, $b$ and $c$ are real due to \eqref{smatrixsplane}.
Next we impose that   $2\delta_{rep} =\frac{s}{4}+O(s^2)$ and 
\beq
\label{imdelta}
\text{Im}\,2\delta_{rep} =\eta_{rep}s^6+O(s^7) \, , 
\eeq
because particle production starts with $|\mathcal{M}_{2\to 4}|^2 \sim  l_s^{12}$.
In fact, the leading term in the probability of particle production 
\beq
P_{rep\to n\ge4\,\text{branons}}=2\eta_{rep}s^6+O(s^7) \, , 
\eeq
is universal and can be computed using the leading order expressions
for $\mathcal{M}_{2\to 4}$ in~\cite{Cooper:2014noa}.
Using the ansatz \eqref{leansatz} in~\eqref{isospinamplitudes} and imposing~\eqref{imdelta} we find
\begin{align}
&2\delta_{sym} =\frac{s}{4}+\a_2 s^2+\a_3
   s^3+\left(\a_4+ \frac{(D-4)    \a_2^2}{2 \pi}\log  s \right)
   s^4\nonumber\\&+\left(\a_5-\frac{(D-4) \a_2 \beta_3 }{\pi }\log s\right)
   s^5\\
  &+ \left(i\eta_{sym} +\a_6+\frac{q \log s }{2\pi(D-2) }\right. \nonumber\\
 &\qquad \qquad \qquad \left.
   +\frac{(D-4)^2 \a_2^3 }{4\pi^2 }\log^2 s\right)
   s^6 +O\left(s^7\right)\nonumber  \,,
      \end{align}
   \begin{align}
   &	2\delta_{anti} =\frac{s}
   {4}-\a_2
   s^2+(\a_3+2
   \beta_3)
   s^3\nonumber\\&-\left(\a_4+\frac{(D-4)  \a_2^2 }{2 \pi }\log s\right)
   s^4\nonumber\\&+\left(\a_5+2
   \beta_5+\frac{D
   \a_2 \beta_3 }{\pi }\log s\right)
   s^5\\
  &+ \left(i\eta_{anti}-\a_6 +\frac{(D-4)\alpha_2^2}{2}\right. \nonumber\\
 &\qquad \left.-\frac{q \log s }{2\pi(D-2) }
   -\frac{(D-4)^2 \a_2^3 }{4\pi^2 }\log^2 s\right)
   s^6+O\left(s^7\right) \nonumber \,,
   \end{align}
and
   \begin{align}
   &	2\delta_{sing} =\frac{s}
   {4}-(D-3) \a_2
   s^2+(\a_3-(D-2)
   \beta_3)
   s^3\nonumber\\&-(D-3)\left( 
  \a_4 +\frac{ (D-4) \a_2^2   }{2 \pi} \log s\right)
    s^4\nonumber\\&+\left(\a_5
-(D-2) \beta_5-\frac{ D(D-3)
  \a_2 \beta_3}{\pi } \log s\right)
   s^5\\
   &+ \left(i\eta_{sing}+\frac{(D-3)\left((D-4)(D-2)\alpha_2^2-12\a_6\right)}{12}\right. \nonumber\\
 & \left.-\frac{q (D{-}3)\log s }{2\pi(D{-}2) }
   -\frac{(D{-}4)^2 (D{-}3)\a_2^3 }{4\pi^2 }\log^2 s\right)
   s^6
    +O\left(s^7\right)\nonumber\,.
\label{fullsinglet}
   \end{align}
where $\a_2,\a_3,\beta_3,\a_4,\a_5,\beta_5,\alpha_6$ are  real parameters and
$q\equiv2\eta_{sing}{-}D \eta_{sym} {+}(D{-}2) (\eta_{anti}{+}2(D{-}4)\a_2\a_4{+}D\beta_3^2 )$.
Notice that the coefficient of the logs do not involve extra free parameters.
The non-linearly realized Poincar\'e symmetry fixes
$\alpha_2=\frac{D-26}{384\pi}$. On the other hand,~$\a_3,\beta_3,\a_4,\a_5,\beta_5,\alpha_6$
are non-universal  parameters related to   $K^4$, $\nabla^2 K^4$, $\nabla^4 K^4$  and $\nabla^6 K^4$ terms in the effective action \eqref{lag}.
The universal coefficient of the $s^4\log s$ terms agrees with the results of~\cite{Conkey:2016qju}. The terms of order $s^5$ and $s^6$ are new.
Notice that  for $D=4$ the first non-analytic term is $s^5\log s$ and it is proportional to the non-universal coefficient $\beta_3$.

\section{Schwarz-Pick}
\label{appendix:SP}
\subsection{Maximum Modulus Principle}
Holomorphic functions are equal to the average of their neighbouring points, 
\beq
f(z) = \oint\limits_{|w-z|=\epsilon} \frac{dw}{2\pi i } \frac{f(w)}{z-w}  = \int\limits_{0}^{2\pi} \frac{ d\theta}{2\pi} f(z+\epsilon e^{i \theta}) \, ,
\eeq
and therefore can not have local maxima or minima inside any domain. All maxima or minima must be at the boundary. 

In particular, if a function's modulus is bounded by $1$ on a boundary of a connected domain and has no singularities inside that domain then the function's modulus must be bounded by $1$ everywhere inside the domain as well. This is known as the maximum modulus principle.

\subsection{Schwarz-Pick}

Schwarz-Pick multi-point lemmas are simple but very powerful extensions of the maximum modulus principle.

Consider a function $f^{(0)}(z)$ regular inside a unit disk and bounded as $|f^{(0)}(z)|_{|z| \le 1}\le1$ everywhere. Out of it construct a new function
\beq
f^{(1)}(z):=\Delta^{(1)}[f](z|w) \equiv \frac{f^{(0)}(z)- f^{(0)}(w) }{1- f^{(0)}(z) \overline{f^{(0)}(w)}} \Big/\frac{z-w}{1-z \overline w} 
\label{SPineq}
\eeq
where $|w|<1$. Since $w$ is strictly inside the disk, this function has no singularities as can be easily checked.
Moreover, it is still bounded along the unit circle: take~$z=e^{i\phi}$, then the combination $(z-w)/(1-\bar w z)$ in eq.~\eqref{SPineq} can be written as
\beq
\frac{z-w}{1-\bar w z}=\frac{1{-}w_r\cos\phi{-}w_i\sin\phi{-}i(w_i\cos\phi{-}w_r\sin\phi)}{1{-}w_r\cos\phi{-}w_i\sin\phi{+}i(w_i\cos\phi{-}w_r\sin\phi)},
\eeq
showing that it is a pure phase (the same happens when we replace~$f^{(0)}(z)$ with $z$).
As such, by maximum modulus principle, it is bounded everywhere and $|f^{(1)}(z)|_{|z| \le 1}\le1$, just like we had for the original function. 

We could now go on constructing a new function $f^{(2)}$ which would depend on a new parameter corresponding to a new point $w^\prime$ strictly inside the unit disk and so on. Each of these new functions would again be bounded everywhere inside the disk. This is the content of the Schwarz-Pick multi-point lemmas~\cite{BeMi}.

There is a cute relation between these lemmas and some $AdS_2$ geometry: 
it is a well established fact in complex analysis that analytic functions are either isometries or contractions of the Poincar\'e disk 
\beq
d^{(h)}(f(z),f(w))\leq d^{(h)}(z,w)\, ,
\eeq
with $d^{(h)}(z,w){=}2\tanh^{-1}|(z{-}w)/(1-z\bar w)|$ the definition of the hyperbolic distance.
Eq.~\eqref{SPineq} is known in the mathematical literature as ``hyperbolic quotient'' and its infinitesimal form
\beq
\frac{d_h f}{d_h z}=f^\prime(z)\frac{1-|z|^2}{1-|f(z)|^2} \, , 
\eeq
is known as the hyperbolic derivative.

The isometries of the disc are given by the single CDD zeros 
\beq
f_{CDD}^{(1)}(z|z_0)=e^{i\alpha}\frac{z-z_0}{1-\bar z_0 z} \, , 
\eeq
and it is easy to check $\Delta^{(1)}[f_{CDD}^{(1)}]$ is a pure phase.
If we consider a generic product of CDD zeros
\beq
f_{CDD}^{(n)}(z|z_0,z_1,\dots, z_{n{-}1})=\prod_{i=0}^{n{-}1}f_{CDD}^{(1)}(z|z_i) \, , 
\eeq
then $\Delta^{(1)}[f_{CDD}^{(n)}] \sim f_{CDD}^{(n{-}1)}$ or equivalently $\Delta^{(n)}[f_{CDD}^{(n)}]$ is a pure phase.
Of course, there are functions like~$e^{(z-1)/(z+1)}$ that, though saturating unitarity, are not CDDs. Indeed they can be represented as infinite products of CDDs 
and they satisfy all the finite $n$ Schwarz-Pick inequalities, but we do not know any functional bound saturated by those functions.

\subsection{Application 1: expansion around the center of the disk}

\begin{figure}[t]
\centering
        \includegraphics[scale=0.28]{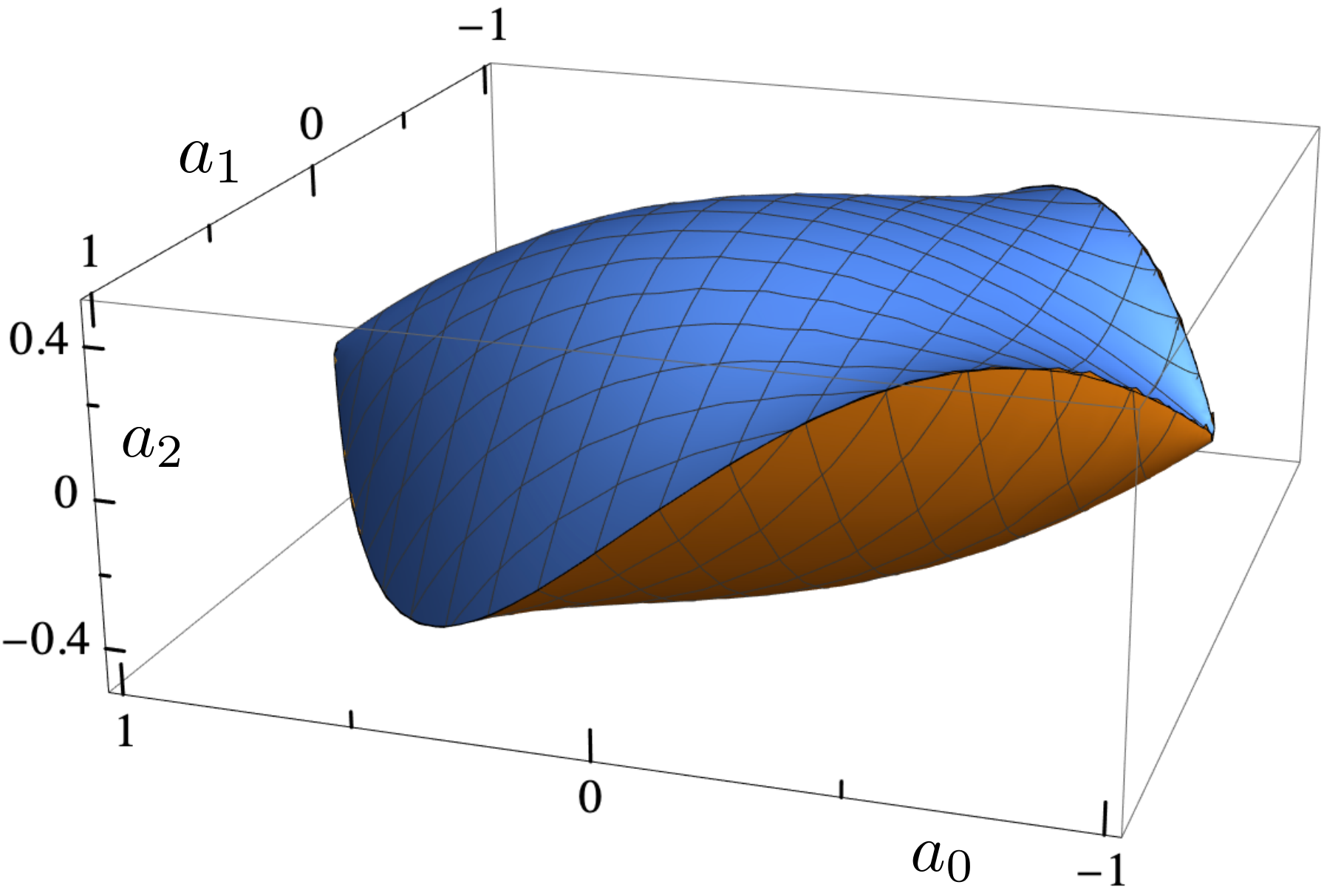}
 \caption{Allowed space for the first three Taylor coefficients $\{a_0,a_1,a_2\}$ of analytic functions from disk to disk as derived using the Schwarz-Pick lemmas.}
\label{derivativemanifold}
\end{figure}

As a first application let us consider a function defined on the unit disk as a Taylor expansion of the form
\beq
f^{(0)}(z)=\sum_{n=0}^\infty a_n z^n,
\eeq
and let us assume it is bounded within the unit disc so~$|f^{(0)}(z)|_{|z|=1}\leq 1$.
The simplest question one could ask is: ``what is its maximum/minimum value for~$z=0$?''.
The answer is given by the maximum modulus principle as explained above and it is $|f^{(0)}|=1$ or equivalently $|a_0|\leq 1$.
A slightly non-trivial question one could then ask is: ``given a value of $a_0$, what are the bounds on the derivative of the function at zero i.e.\ $a_1$?''
If we construct $\Delta^{(1)}[f^{(0)}]$, or alternatively the hyperbolic derivative and we require its boundeness we readily get that~$|a_1|\leq 1{-}a_0^2$.
It is easy to guess that applying $\Delta^{(2)}$ will give a bound on $a_2$ as a function of $a_0$ and $a_1$
\beq
|a_2|\leq \frac{(a_0^2-1)^2+a_1^2(2 a_0-1)}{2(a_0^2-1)} \, , 
\eeq
yielding the nice manifold in figure~\ref{derivativemanifold}.
We cannot plot the higher order constraints but we can derive them analytically. What we would get is an algebraic manifold contained in the vector space
of the Taylor expansion coefficients where all the analytic functions on the unit disc reside.
In our language, this is the space of generic massless S-matrices in $1+1$ dimensions.

\subsection{Application 2: expansion around the threshold}
\label{SPapplication}

Let us now turn to the application of the above inequalities to $f^{(0)}$ the $D= 3$ flux-tube branon S-matrix when $z,w,w^\prime,{\dots}$ are all close to $1$. When translated back to $s$, this limit corresponds to the very low energy region where we can relate the coefficients of the Taylor expansion of~$f^{(0)}$ to the effective field theory parameters. 
Furthermore, we assume that the threshold is a regular point, which is not the most general behavior for flux tube S-matrices, but, as explained in appendix~\ref{ap:deltaexpansions}, it is true up to some high order in $s$.

For instance, if
\beq
f^{(0)}=S=e^{i(\gamma_1 s+i \gamma_2 s^2+\gamma_3 s^3 +i \gamma_4 s^4+\gamma_5 s^5+\dots)} \, , 
\label{3dgeneric}
\eeq
and we take $s=\epsilon e^{i \theta}$, with $-\pi\leq\theta\leq \pi$, then the maximum modulus principle implies that
\beq
|f^{(0)}|{=}1{-}2\sin\theta \gamma_1 \epsilon {+}(2 \sin^2\theta\gamma_1^2{-}2\cos{2\theta}\gamma_2)\epsilon^2{+}\mathcal{O}(\epsilon^3)\leq 1.
\eeq
For generic $\theta$ this condition is equivalent to a positive condition on $\gamma_1\geq 0$.
Since $\gamma_1$ is related to the strength of the tree-level interaction, we notice that in this simple framework the maximum modulus principle is equivalent to the causality bound derived in~\cite{Adams:2006sv}.
Moreover, at~$\theta = 0$ we don't get any bound on $\gamma_1$, but only on $\gamma_2\geq 0$. Notice that the maximum modulus principle for $\theta = 0$ is equivalent to unitarity.
This is a general feature of the parametrization we chose in eq.~\eqref{3dgeneric} compatible with crossing and real analyticity: odd powers are pure phases, even powers contribute to inelasticity.
Therefore, we could set $\gamma_{2n}= 0$ and focus only on the pure phase coefficients where unitarity has nothing to say about them.

\begin{figure}[t]
\centering
        \includegraphics[scale=0.24]{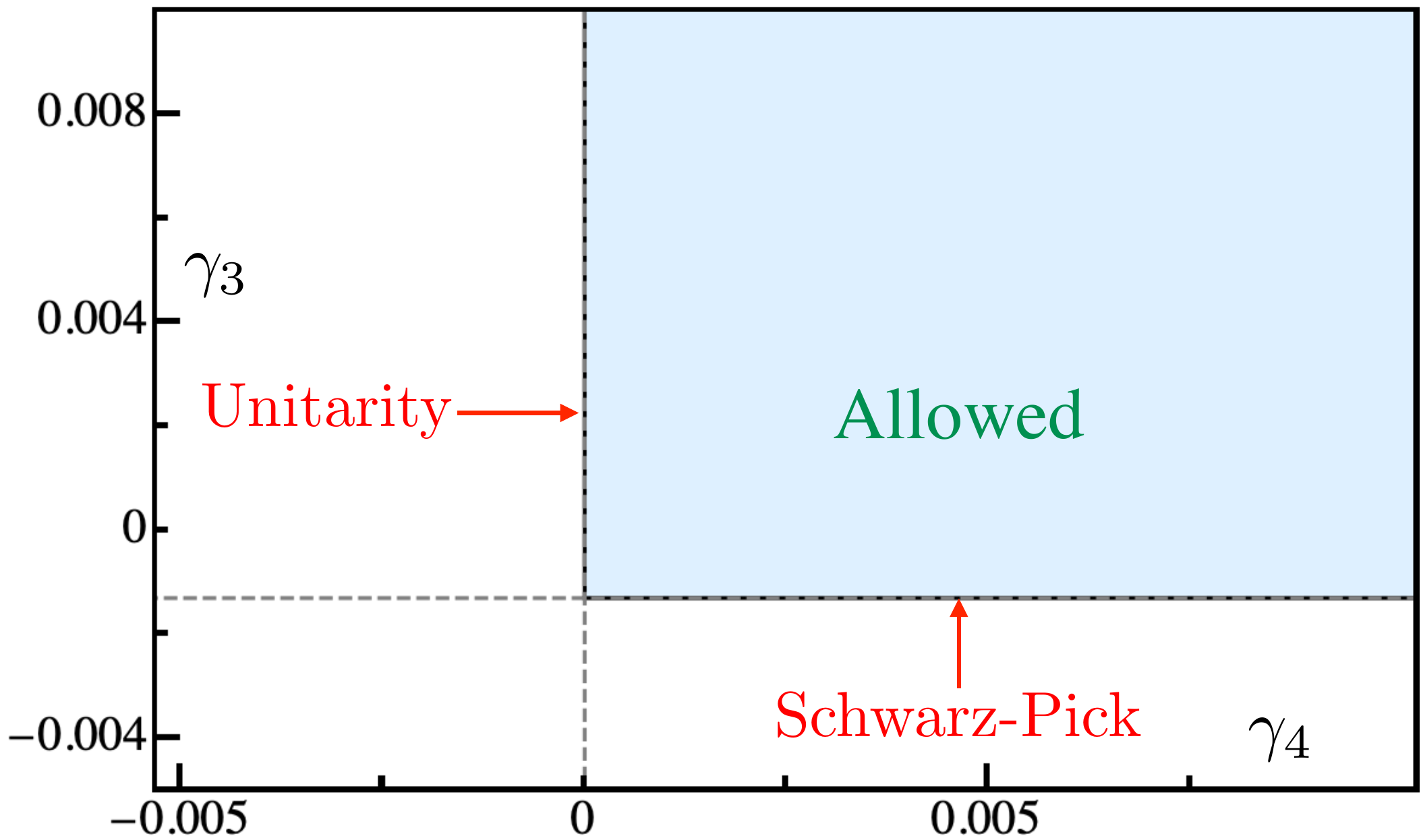}
 \caption{ Allowed region in the $\{\gamma_3,\gamma_4\}$ plane. The unitarity
vertical bound is saturated by the integrable S-matrix family in~\eqref{CDDedge}.
 The Schwarz-Pick bound would be saturated
by theories with a very small amount of particle production localized at threshold for which we don't have an analytic ansatz. The corner at the intersection of the two bounds is saturated by a known integrable theory describing the RG flow from tricritical Ising fixed point to the free fermion. } 
\label{3Dfigureold}
\end{figure}

To make contact with flux tube theories~\eqref{lee3D} 
 we set~$\gamma_1 =1/4$.
We map the upper half-plane to the unit disk and taking first $z\to w$ and then $1{-}z = \eps e^{i\phi}{+}\mathcal{O}(\epsilon^2)$ in the Schwarz-Pick lemma 
we are left with
\beq
1 {\ge} |f^{(1)}| = 1{-}16 \sec\phi\,\epsilon \gamma_2+\mathcal{O}(\epsilon^2){+}{\dots}
\eeq
If unitarity is saturated the first term vanishes. At the next order 
\beq
|f^{(1)}| =1-\frac{1}{12}\epsilon^2 \cos^2\phi (1+768 \gamma_3)+\mathcal{O}(\epsilon^3) \, , 
\eeq
the term of $\mathcal{O}(\epsilon^2)$ must be negative leading to 
\beq
\gamma_3\geq -\frac{1}{768}.
\eeq
Combining the constraint above with the one from unitarity ~$\gamma_4\geq 0$ at order $\epsilon^4$ we get figure~\ref{3Dfigureold}.
Assuming unitarity saturation, i.e.\ $\gamma_{2n} =0$, we can recursively generate new constraints on the higher derivative coefficients using higher order multi-point lemmas
-- see also~(\ref{rockinequalities}) and figure~\ref{3Dfigure}.

If at some order inelasticity kicks in then the Schwarz-Pick inequalities are not saturated and we no longer get any sharp bounds. 

\begin{figure}[t]
\centering
        \includegraphics[scale=0.49]{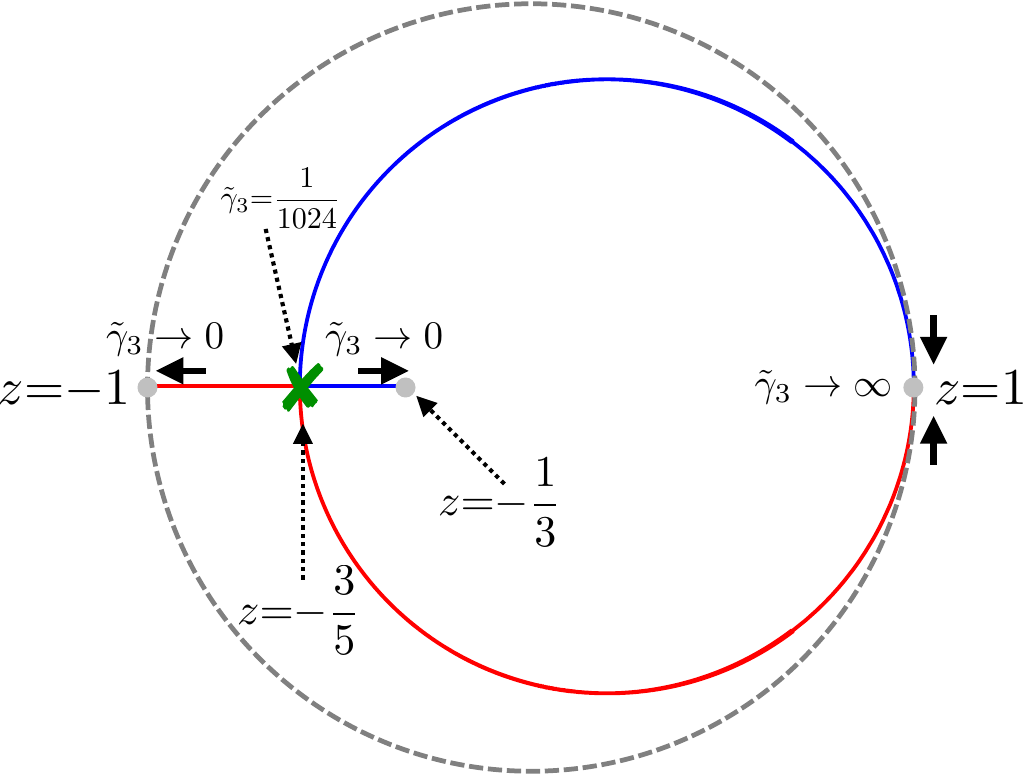}
 \caption{ Resonance positions as we move along the red edge of figure 1: At the cusp we have a virtual state at $\chi={-}1/3$ (or $s= 8 i$) and the S-matrix can be identified with the goldstino S-matrix of Zamolodchikov~\cite{Zamolodchikov:1991vx}. As we move away from the cusp another zero comes in from $i\infty$; eventually the two zeros collide at $s=16 i$ and move away acquiring a real part; in this region they become closer and closer to more conventional sharp resonances until they eventually collide with the $s= t = 0$ threshold.} 
\label{movie}
\end{figure}

As explained in the main text, figure~\ref{3Dfigure} represents the space of all integrable S-matrices compatible with the~$D=3$ flux-tube universal low-energy behaviour~\ref{lee3D}.
Using the language introduced in this section we would interpret its geometric features saying that
the cusp saturates $\Delta^{(1)}$, the red edge $\Delta^{(2)}$ and the surface $\Delta^{(3)}$ Schwarz-Pick constraints. 
For this reason, it is a theorem that the cusp S-matrix is given by a single CDD, the red edge by products of two CDD and the orange surface by products of three CDD factors
\begin{align}
&S_{\text{surf}}=\nonumber\\
&\frac{(1{+}3 z)(5{+}z(6{+}5 z))\tilde\gamma_3{-}256(z{-}1)^3\tilde\gamma_3^2{+}64\tilde\gamma_5(z{-}1)^2(3z{+}1)}{(z{+}3)(5{+}z(6{+}5 z))\tilde\gamma_3{+}256(z{-}1)^3\tilde\gamma_3^2{+}64\tilde\gamma_5(z{-}1)^2(z{+}3)} \la{surf}
\end{align}
which for $\tilde\gamma_3=0$ and $\tilde \gamma_5=0$ reduces to
\beq
S_{\text{cusp}}=\frac{1+3z}{z+3},
\eeq
and saturating the second Scwharz-Pick inequality $\tilde \gamma_5=4\tilde\gamma_3^2-1/64\tilde\gamma_3$ reduces to
\beq
S_{\text{edge}}=\frac{1+4 z+3 z^2+128(z-1)^2\tilde\gamma_3}{3+4 z+z^2+128(z-1)^2\tilde\gamma_3}\,.
\label{CDDedge}
\eeq
whose resonance positions are shown in figure~\ref{movie} in the unit disk (to go to the $s$ plane, use $z=\chi(s)=(4+is)/(4-is))$.

Note that these resonances along the red edge of figure~\ref{3Dfigure} can be separated into two possibilities -- see also figure~\ref{movie}: 
they can both be at purely imaginary $s$ (or real $\chi$) or they can be in a pair, symmetric with respect to reflections on the imaginary axis (i.e.\ given by a complex conjugate pair in~$\chi$). These two possibilities are separated by a ``collision", represented by the green cross in figure~\ref{movie}. At that point, the two zeros collide and the S-matrix simply becomes a perfect square with a double zero. This happens at some point along the red edge of the three dimensional figure~\ref{3Dfigure}. 

\begin{figure}[t]
\centering
        \includegraphics[scale=0.6]{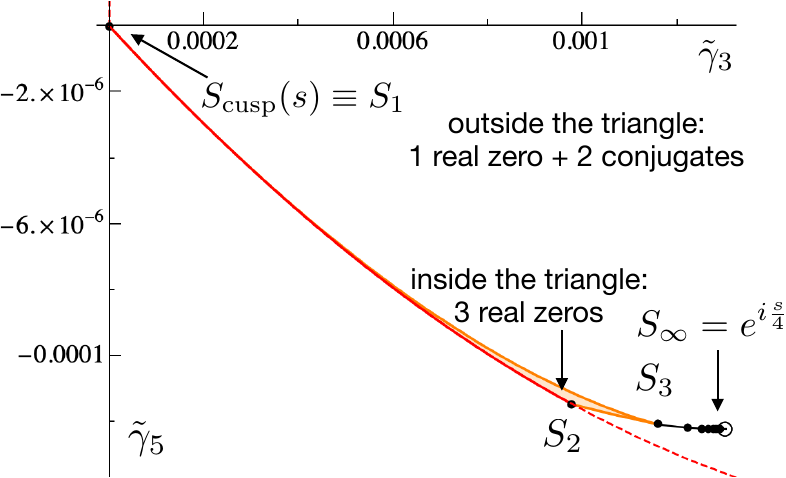}
 \caption{The triangular area depicted in orange represents the face in the allowed $\{\tilde\gamma_3,\tilde\gamma_5\}$ region where the three zeros of the S-matrix are real. The three edges of the triangle correspond, respectively, to one single CDD $S_{1}$, a double CDD $S_{2}$, and a triple CDD zero $S_{3}$. The red edge between $S_{1}$ and $S_{2}$ is the projection of a finite arc of the red edge in fig.~\ref{3Dfigure} and correspond to two real distinct CDD zeros. All the S-matrices on the orange edges contain at the same time a double and a single CDD zero. 
 Finally, the interior of the triangle has three CDD zeros all distinct. The unlabeled black dots are the projections of the higher order zeros in~\eqref{Sn} which converge to $e^{is/4}$.
 } 
\label{tetraA}
\end{figure}

\begin{figure}[t]
\centering
        \includegraphics[scale=0.6]{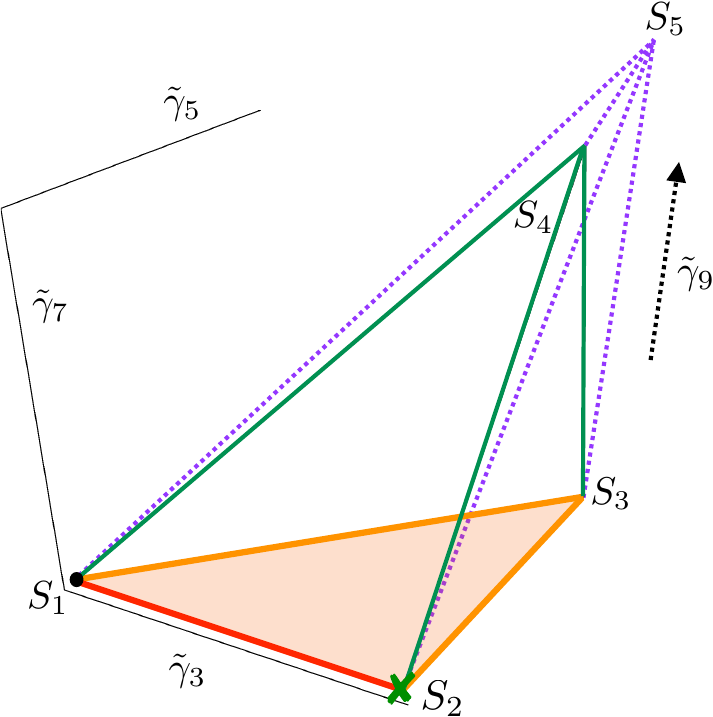}
 \caption{ Schematic representation of the compact region in the allowed $\{\tilde\gamma_3,\tilde\gamma_5,\tilde\gamma_7\}$ space where all the zeros are purely real. In the interior of the tetrahedron there are 4 distinct CDD zeros. 
 As we hit one of the faces two things can happen: two zeros collide becoming a double zero or, as in the case of the orange face (the same shown in figure~\ref{tetraA}), one zero goes at infinity. 
 The purple dotted lines are the three-dimensional projections of the edges of the hyper-tetrahedron in the allowed $\{\tilde\gamma_3,\tilde\gamma_5,\tilde\gamma_7,\tilde\gamma_9\}$ space: every time we add an extra dimension a higher multiplicity zero cusp appears. Extrapolating to $n\to \infty$ we encounter the $e^{is/4}$ cusp.
 } 
\label{tetraB}
\end{figure}

In the same figure we have the S-matrices on the orange surface, given by (\ref{surf}) which have three zeros and again, there is a point on that surface where the three collide. The corresponding $S$-matrix at that point has a triple zero.
Together with the black cusp in figure \ref{3Dfigure} which was given by a single CDD factor, we see that we have a family of three S-matrices
\beq
S_{n} = (-1)^n \left(\frac{s-8 i \,n}{s+8 i \,n}\right)^n
\label{Sn}
\eeq
with $n=1,2,3$ which we can single out as somehow special. We can connect them by a triangle as represented in figure~\ref{tetraA}. 
The region inside the triangle is compact, in contrast to the infinite orange surface in figure~\ref{3Dfigure}; 
it is defined by the condition that all zeros there are purely real (in $\chi$). 
Again, let us stress that $n=1$ is related to a cusp, $n=2$ to an edge and $n=3$ to a face.

There is a clear higher dimensional generalization we could make here if we stick to the family of purely elastic  $S$-matrices that saturate unitarity and therefore have   $\gamma_{2n}=0$.
If we further add $\gamma_9$ to our analysis, for example, we would now have a four dimensional space and the triangle would be the base of a tetrahedron with the fourth new vertex corresponding to an $S$-matrix $S_{4}$ corresponding to the collision of four zeros. Inside the tetrahedron all zeros would be real (in $\chi$). This is schematically represented in figure~\ref{tetraB}. The tetrahedron itself would be the base for a 4 dimensional polyhedra with an extra cusp $S_{5}$ 
(schematically, this is where the dashed purple lines in the figure would meet -- of course we can only draw their three-dimensional projections). 
This would go on thus defining a sequence of $S$-matrices given by~\eqref{Sn} for any $n\ge  1$. 

So we could ask whether this sequence of vertices in this infinite dimensional space of polyhedra would converge towards anything interesting. Indeed, beautifully, we have 
\beq
\lim_{n\to \infty} S_{n} = e^{i s/4} \,
\eeq
the famous integrable flux tube S-matrix~\cite{Dubovsky:2012sh,Dubovsky:2012wk}. In practice, already for $n=3$ on the orange surface we would be very close to $e^{i s/4}$ for most values of $s$. At higher energies we would see deviations. The higher $n$ is in~\eqref{Sn}, the larger is the range in $s$ where the S-matrix is indistinguishable from the integrable flux tube S-matrix.

\section{Yang-Baxter equation and analytic solution}\label{YBAppendix}
For massless S-matrices with $O(N)$ symmetry, the Yang-Baxter equation takes the form:
\beq \label{yb}
S^{c b_2}_{a_2 a_3}(\theta_{23}) S^{b_3 b_1}_{a_1 c} (\theta_{13}) = S^{c b_1}_{a_1 a_3}(\theta_{13}) S_{a_2 c}^{b_3 b_2} (\theta_{23}) \, , 
\eeq
where $\theta$ is the rapidity defined by $p_L = -\exp(-\theta)$ for left-movers and $p_R = \exp(\theta)$ for right-movers and $\theta_{ij} \equiv \theta_i - \theta_j$. 

In terms of the amplitudes defined by the equation $\ref{4dsmatrixdecomp}$, this implies the condition $\sigma_1 = \sigma_3 = 0$ for $N > 2$. However for $N=2$, we have the following relaxed condition: 
\beq \label{ybsigma}
\sigma_3(s) = -\sigma_1(s) \, . 
\eeq 
This is the case for flux tubes in 4 dimensions, where the remnant symmetry of the goldstone bosons is $O(2)$. In terms of the isospin amplitudes, this implies that~$\sigma_{sing} = \sigma_{anti} = \sigma_1 + \sigma_2$ and $\sigma_{sym} =  \sigma_2 - \sigma_1$. Moreover, we can make use of the crossing and analyticity relations~\eqref{smatrixsplane} to deduce that
\beq \label{sigmaintegrability}
\sigma_{sym}(-s^*) = \sigma_{sing}(s)^* = \sigma_{anti}(s)^*  \, . 
\eeq
The symmetric channel S-matrix is bounded by~$1$ for any $s \in \mathbb{R} $, hence ~\eqref{sigmaintegrability} now implies that the antisymmetric and the singlet channel amplitudes are also bounded by~$1$ for $s \in \mathbb{R}$. Under the assumption of integrability, these inequalities are saturated and the isospin amplitudes are a product of CDDs.\footnote{Any holomorphic function $f(z)$ from the upper half plane $\mathbb H$ to the unit disc $\mathbb D$ that satisfies $|f(z)| = 1$ for $z \in \mathbb R$ must be a product of CDDs.} We therefore consider the following simple ansatz for the solution that satisfies Yang-Baxter:
\begin{eqnarray}
\sigma_{sym}&=& \frac{(s-a)a^*}{(s-a^*)a} \nonumber \\
\sigma_{sing}=\sigma_{anti} &=& \frac{(s+a^*)a}{(s+a)a^*}
\label{YBsolution1}
\end{eqnarray}
where we have used (\ref{sigmaintegrability}) to relate the CDD zeroes in the 3 channels. 
To find the location of the zero $a$, we expand the above ansatz at $s= 0$ and match with the low energy expansion \eqref{lowenergyexpansion} and we find $a=8/(32 \alpha_2- i)$. 
We can also read of the following relation between the~$\alpha_2$ and $\alpha_3$ coefficients
\beq
\alpha_3 = -\frac{1}{768} + 4 \alpha_2^2 \, . 
\label{SPGeneric}
\eeq
Substituting   $\alpha_2 = -\frac{22}{384 \pi}$, we get $a\approx -3.5+i 6.0$ and
\beq
\alpha_3 = -\frac{1}{768} + \frac{121}{9216 \pi^2}\,, 
\eeq
which saturates the Schwarz-Pick bound (\ref{SPSym}). 
\section{Numerics}
\label{ap:numerics}

For numerics we parametrize smooth S-matrices as Taylor expansions on the upper half plane or -- mapped to the unit disk -- as 
\begin{eqnarray}\label{chiseries}
\sigma_{1}(\chi) &= &\sum_{n=0}^{N_{max}} (a_n + i b_n) \chi^n\nonumber \\
\sigma_{2}(\chi) &=& \sum_{n=0}^{N_{max}} c_n \chi^n\nonumber\\
\sigma_{3}(\chi) &=& \sum_{n=0}^{N_{max}} (a_n - i b_n) \chi^n 
\end{eqnarray}
where real analyticity simply amounts to the statement that the coefficients $\{a_n,b_n,c_n\}$ are real and the $\chi$ map was defined in figure~\ref{unitDisk}. Given a set $\{a_n,b_n,c_n\}$ we can simply expand the amplitudes close to $s= 0$ (or $\chi =1$) to read off the threshold parameters~\eqref{lowenergyexpansion}. 

As mentioned in the main text, the isospin amplitudes diagonalise unitarity constraints and in the $\chi$ disc we have: 
\beq \label{unitaritychi}
|\sigma_{rep} (e^{i \theta})| \leq 1 \, ,  \quad  \forall \; \theta  \in [0,\pi] \, . 
\eeq
This is easily imposed as a semi definite constraint: We first divide the range of $\theta$ into a grid, and then for a given point on the grid and for each one of the isospin amplitudes, eq.~\eqref{unitaritychi} is the same as the condition
\beq
\mathcal U \equiv \begin{pmatrix}
1+\mathcal R &  \mathcal I\\
\mathcal I & 1-\mathcal R \\
\end{pmatrix} \succeq 0 \, . 
\eeq
where $\mathcal R = \Re(\sigma_{rep}(e^{i \theta}))$ and $\mathcal I = \Im(\sigma_{rep}(e^{i \theta}))$.

Experimentally, we find that a Chebyshev grid of points gives the best results and a grid size of around~$200$ points is sufficient for $N_{max}$ all the way up to 100. 

\subsection{The space of S-matrices compatible with $D=4$ flux tubes universality} 

In section~\ref{D4section} we fixed $\alpha_2$ to be the universal value predicted by non-linearly realized Poincar\'e as for flux tube theories.
From a general S-matrix perspective it is a legitimate question to first ask about the allowed space of~$\{\alpha_2,\beta_3,\alpha_3\}$ parameters and look for any structure 
pointing to the physical section at fixed~$\alpha_2$.
So, the question we ask is: ``what is the minimum of $\alpha_3$ at fixed $\alpha_2$ and~$\beta_3$?''

\begin{figure}[t]
\centering
        \includegraphics[scale=0.29]{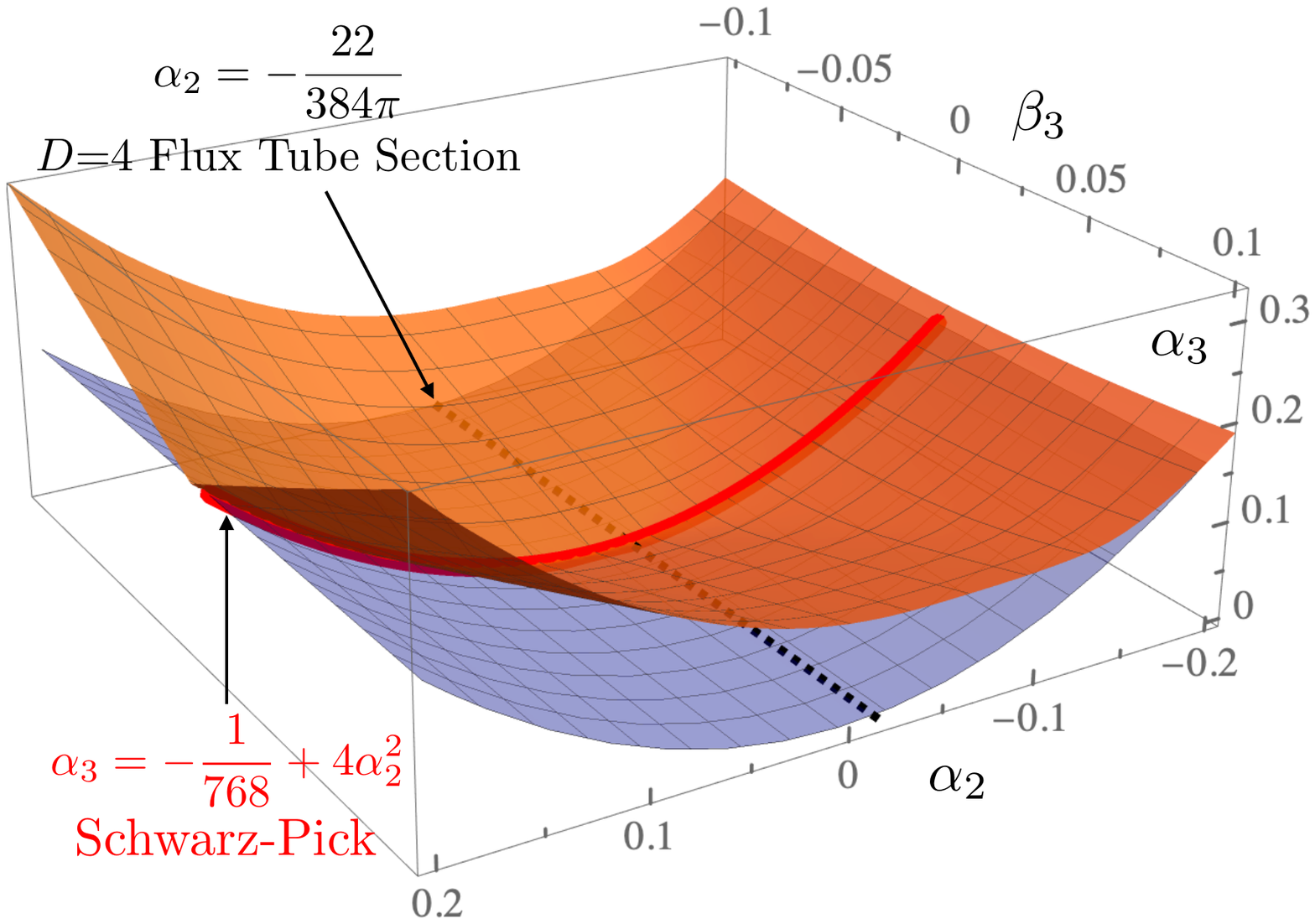}
 \caption{Allowed region in the generic $\{\alpha_2,\beta_3,\alpha_3\}$ parameter space compatible with $D=4$ flux tube S-matrices. The light blue surface is given by the Schwarz-Pick bound in the symmetric channel; the orange surface by the numerics. They are tangent along the line $\beta_3=0$ (in red) where the S-matrix satisfy Yang-Baxter and is given in appendix~\ref{YBAppendix}. The black dashed line denotes the flux-tube  $\alpha_2$ as predicted by the 1-loop universal $2\to 2$ scattering.}
\label{boundarybound}
\end{figure}

The answer is shown in figure~\ref{boundarybound}. The orange surface is the numerical minimum bound, the blue surface is the Schwarz-Pick analytic bound~\eqref{SPGeneric} and the red line is their intersection.
The black dashed line denotes the physical section at fixed $\alpha_2=-\frac{22}{384\pi}$. There is no sign along this general boundary that any value of $\alpha_2$ plays a special role, except perhaps, $\alpha_2=0$: it seems that at fixed $\alpha_2>0$ there is a cusp for $\beta_3=0$ that becomes smooth as we go to $\alpha_2<0$. We do not know the reason for this, since we did not fully explore the features of the S-matrices saturating this minimal surface in general. It would be nice to perform a detailed analysis in the future.

The red line at $\beta_3=0$, as explained in Appendix~\ref{YBAppendix}, is saturated by S-matrices satisfying Yang-Baxter and we have analytic solutions for them, see eq.~\eqref{YBsolution1}. 

\subsection{Resonances along the boundary at the physical values of $\alpha_2$.}

\begin{figure}[t]
\centering
        \includegraphics[scale=0.40]{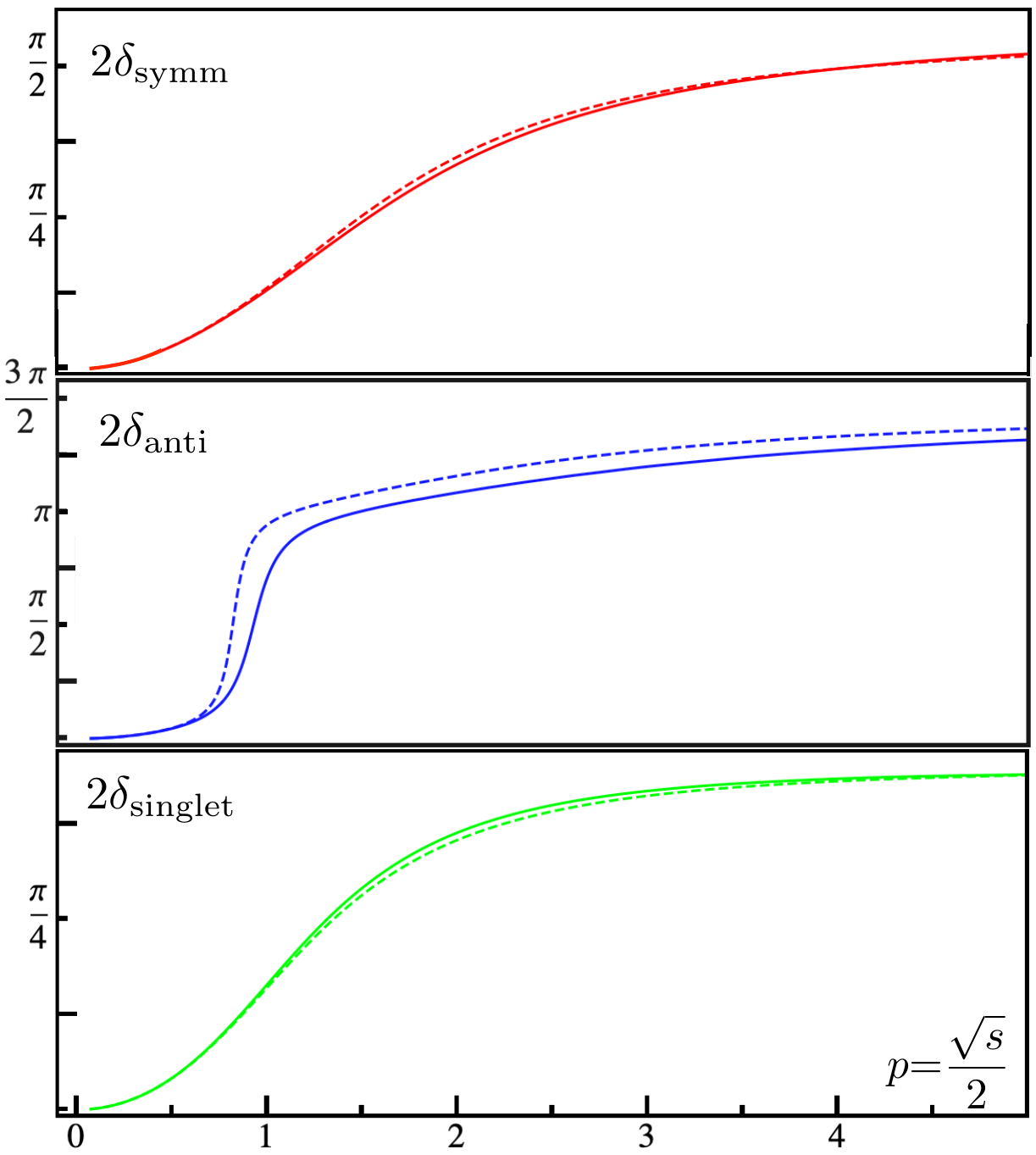}
 \caption{The phase shifts respectively in the symmetric, antisymmetric and singlet channel at the $SU(3)$ point (solid line) and $SU(5)$ point (dashed line).}
\label{figphaseshifts}
\end{figure}

In figure~\ref{figphaseshifts} we plot the phase of each irrep channel as a function of the momenta -- in solid/dashed the $SU(3)/SU(5)$ point S-matrix.
These graphs show nice phase-shifts that are characteristic of resonance behaviour. 
 Such phase-shifts are generated by zeros in the complex $s$-plane,   see ref.~\cite{Doroud:2018szp} for a discussion.
 Measuring zeros in the complex energy plane from experimental or lattice MC data  is not an easy task and in general one needs to use dispersive methods to analytically continue the  
real data to the complex plane; a procedure often plagued by numerical instabilities.
Fortunately, we have the full S-matrix in the physical upper half $s$-plane (or the $\chi$ unit disk) and thus we can easily identify the zeros corresponding to the phase-shifts of   figure~\ref{figphaseshifts}.

\begin{figure}[t]
\centering
        \includegraphics[scale=0.36]{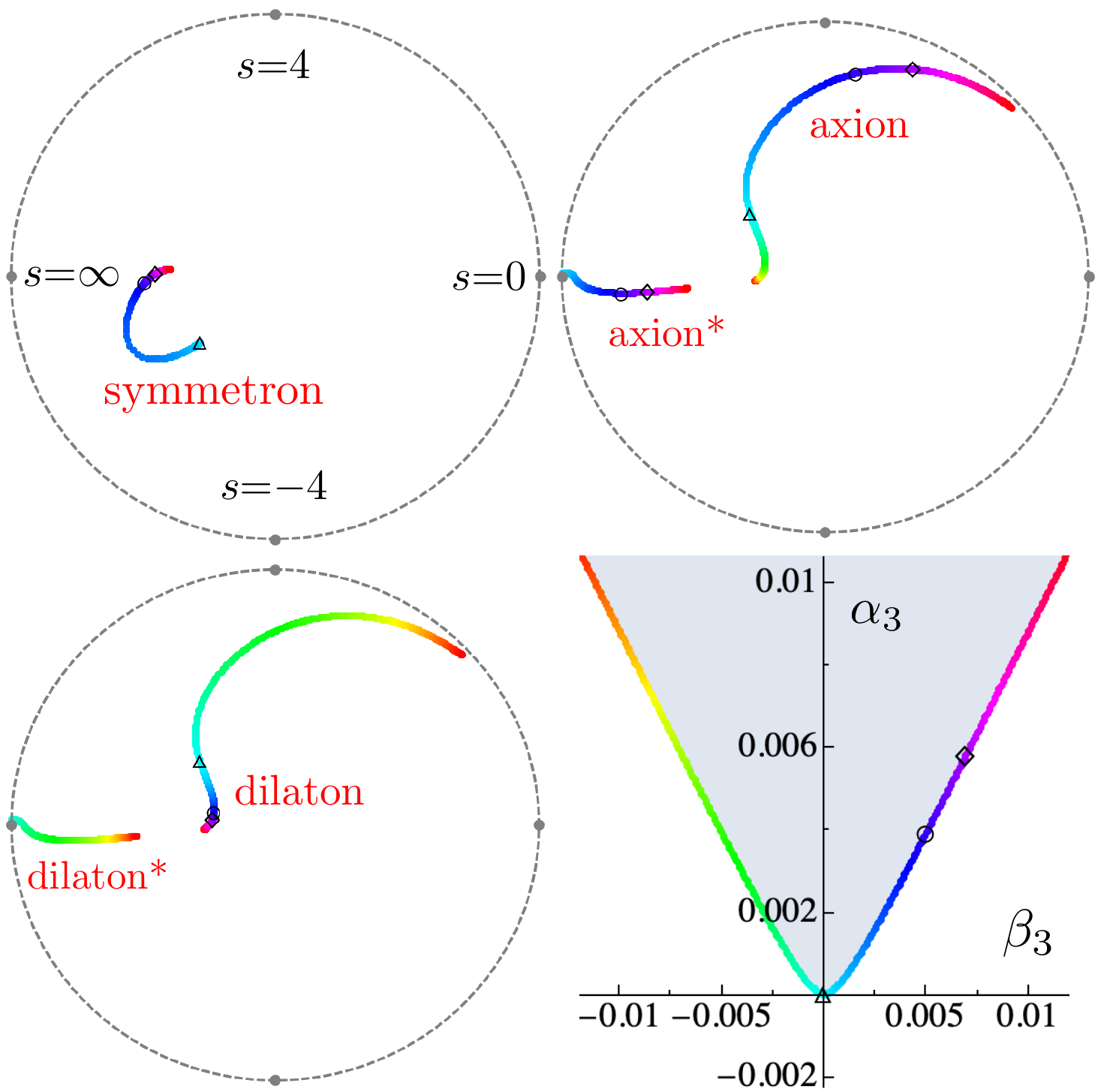}
 \caption{ Zeros trajectory in the unit disk as we move along the boundary of the allowed region in the $\{\alpha_3,\beta_3\}$ space (bottom-right) plot. In each channel we mark with an upper-triangle the zeros at the integrable point, with a circle those at the $SU(3)$ point and with a diamond at the $SU(5)$ point.}
\label{figzeros}
\end{figure}

For  the flux tube value $\alpha_2=-\frac{22}{384\pi}$, there is a unique S-matrix at each boundary point of the  $\{\beta_3,\alpha_3\}$ space   shown in figure~\ref{fig4dboundary}. 
 In figure~\ref{figzeros} we show the position of the resonances  for the S-matrices in a section of the boundary. 
The $SU(3)/SU(5)$ benchmark points in the table of sec.~\ref{spectrum} are denoted with a circle/diamond.
The triangle signals the resonances at the integrable point $\beta_3=0$. 
Note that the symmetry $\beta_3\leftrightarrow -\beta_3$ of the crossing equations  introduced in the main text
is now visible in the symmetric positions of the resonances in the singlet and antisymmetric channels. 
The symmetron resonance is invariant under $\beta_3\leftrightarrow -\beta_3$, hence we only show $\beta_3>0$.

\subsection{Exploring the boundary and the spectrum fixing the axion} \la{axionAppendix}

\begin{figure}[t]
\centering
        \includegraphics[scale=0.33]{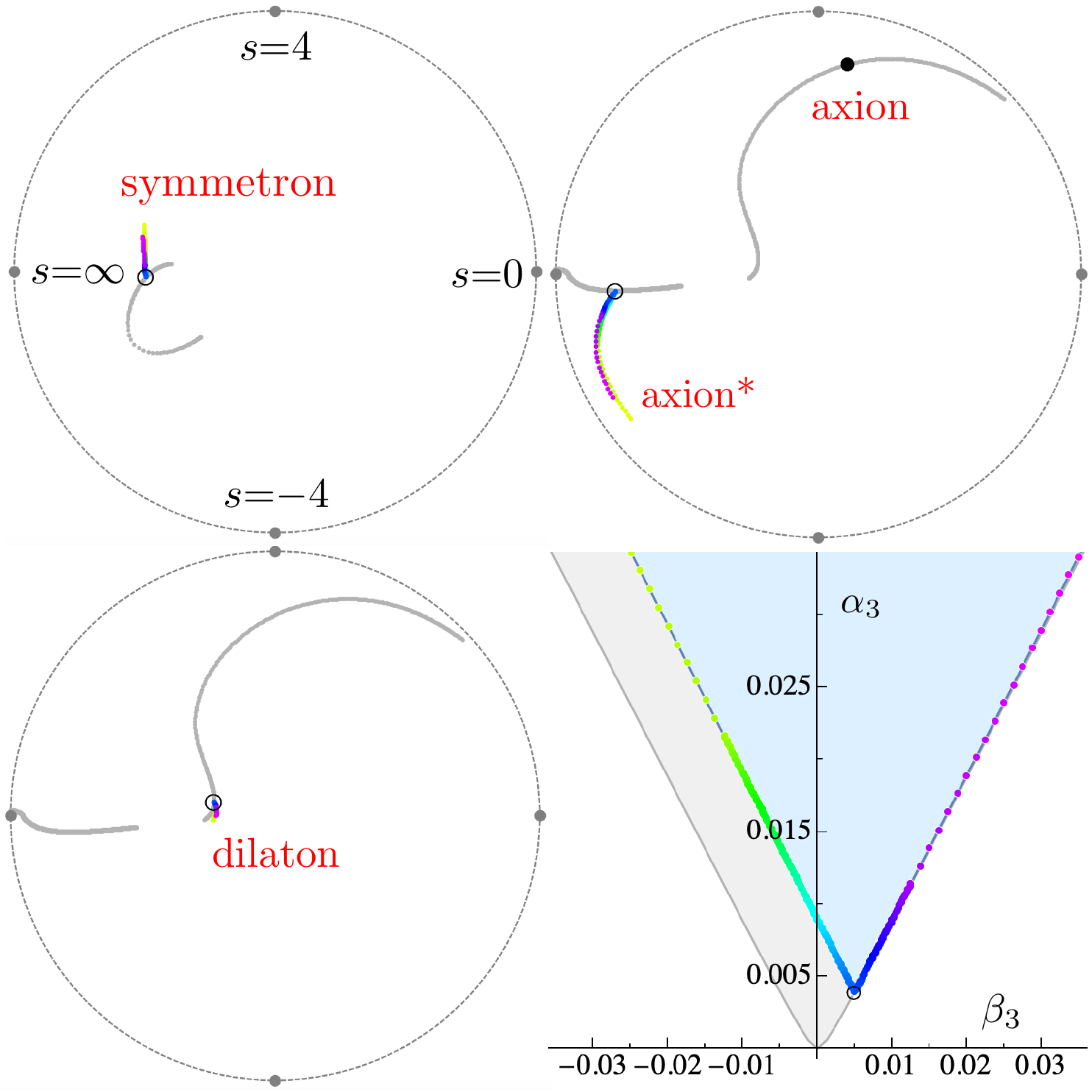}
 \caption{Zeros trajectory in the unit disk of each irrep S-matrix as we move along the boundary of the allowed region in $\{\beta_3,\alpha_3\}$ parameter space (bottom-right) \textit{at fixed} $SU(3)$ axion as given in tab.~\ref{spectrum}. In gray are the old resonance trajectories without the axion imposed. In the bound figure (bottom-right) we show in gray shades the old smooth boundary: the presence of a fixed resonance sharply cut the allowed region giving rise to a kink at the point where the optimal bound allows to emerge the resonance we fix.}
\label{axionfixed}
\end{figure}

The last numerical problem we address in this section is what happens to the spectrum if we fix the experimentally observed world-sheet axion.
For instance, we can minimize the value of $\alpha_3$ at any fixed $\beta_3$ and look at the S-matrix optimizing the bound, given the additional condition that $S(s_{\text{axion}}) = 0$.
The result when we fix the axion at the $SU(3)$ value is shown in figure~\ref{axionfixed}.

As we might expect, imposing an additional condition shrinks the allowed space of parameters, but interestingly, while the previous bound was smooth, now there is a kink.
Moreover, the resonance spectrum is somehow stable: for any value of $\beta_3$ we find an axion*, a dilaton and a symmetron and for the range of $\beta_3$ we scanned their position does not vary much. Another game one could play is to fix $\beta_3$, $\alpha_3$ and the axion at the experimentally estimated values and repeat the analysis of the spectrum varying some other higher order parameter or
bounding it. We leave this interesting analysis to future explorations.

\subsection{Coupling $Q$} \label{resonancesAppendix}

When a resonance is close  to the real energy axis, the phase of the S-matrix  jumps by~$\pi$ as it passes close to it.
A neat example is  given in figure~\ref{figphaseshifts} where there is a clear jump in the phase of the antisymmetric channel when evaluated close to the resonance. To estimate the coupling to the sharp axion resonance we do a  ``narrow width'' approximation
\beq
S(s)\big |_{s\sim m^2_{\text{res}}}= -\frac{s+s_m-i s_\Gamma/2}{s-s_m+is_\Gamma/2}=e^{2i  \delta_{\text{res}}(s)} \, ,  \label{narw1}
\eeq
where the unitarity cuts are neglected and $s_\Gamma = m_{\text{res}}\Gamma$. Therefore,  
\beq
\frac{1}{\Gamma}=\sqrt{s}\,\frac{\partial \delta_{\text{res}}(s)}{\partial s}\bigg|_{s=m^2_{\text{res}}} \, ,  \label{gammadef}
\eeq
up to terms of $\mathcal{O}(s_\Gamma/s_m)$.
From equation \rf{gammadef} and the parametrisation of the phase   as a function of $Q$ 
 $
2 \delta_{\text{res}}(s)|{=}\arctan{\frac{Q^2 s^3}{8(m^2-s)}}
$ ~\cite{Dubovsky:2014fma, Dubovsky:2015zey},
we have  
$
\Gamma={m^5Q^2}/{8} \, . 
$
which is used  in section~\ref{section:resonances}.

\section{Integrability computations}
\subsection{TBA}
\label{appendixtba}
When expanding the TBA equation at large radius as discussed in the text we end up with simple integrals to evaluate. In fact, all we need is 
\beq
\int \frac{x^n}{e^x-1}=n!\,\zeta_{n{+}1} \equiv I_{n} \la{integrals} \, ,
\eeq
plus two other integrals obtained from this one by integrating by parts:
\beqa
\int\! x^n\log(1-e^{-x}){=}-\frac{I_{n+1}}{n+1}\, ,\, \int\!\frac{x^n}{(e^x-1)^2}{=}-n I_{n-1} {-}I_n \,. \nn
\eeqa
Only even zeta's are generated leading to all the $\pi$'s in the final result (\ref{finalE}). Expanding to higher orders we find 
\beqa
E_0&=&\sqrt{R^2-\tfrac{\pi}{3}}-\frac{32 \pi ^6 \gamma _3}{225 R^7}
-\frac{64 \pi ^7 \gamma _3}{675 R^9}-\frac{\frac{2 \pi ^8 \gamma _3}{45} +\frac{32768 \pi ^{10} \gamma
   _5}{3969}}{R^{11}} \nn \\
   &-&\frac{\frac{16384 \pi ^{11}
   \gamma _3^2}{4725}+\frac{22 \pi ^9 \gamma _3}{1215}+\frac{32768 \pi ^{11} \gamma
   _5}{3969}}{R^{13}} \\
   &-&\frac{\frac{208384 \pi ^{12} \gamma _3^2}{50625}+\frac{1001 \pi ^{10} \gamma _3}{145800}+\frac{26624 \pi
   ^{12} \gamma _5}{5103}+\frac{524288 \pi ^{14} \gamma _7}{225}}{R^{15}} \nn \\
   &+&O(R^{-17}) \,. \nn
\eeqa

For $D=4$ we have two goldstone particles and hence two pseudo-energies and thus a priori, we have two coupled differential equations to solve. Nicely, for the ground state energy they can be reduced to a single equation \cite{ Dubovsky:2014fma} which differs from the $D=3$ equations~\eqref{tba},~\eqref{Etba} in a few factors of $2$ only, precisely stemming from the fact that we have now twice as many goldstone particles:
\beqa
{E}_0(R)&=&R+\frac{{\color{red}2}}{\pi R}\int_0^\infty dx\,\log(1-e^{-\varepsilon(x)})
\label{Etba4D} \\
\varepsilon(x)&=&x+\frac{1}{2\pi} \int_0^\infty \frac{dx^\prime}{x^\prime}\mathcal{K}
\log(1-e^{-\varepsilon(x^\prime)}),
\label{tba4D} \\
\mathcal{K}&=& x' \frac{\partial ( \delta_{sym}{+} \delta_{anti})}{\partial x'} = {\color{red} 2} \Big( \frac{x x'}{R^2}{+}3 \alpha_3 \Big(\frac{4 x x'}{R^2}\Big)^3{+}{\dots}\Big) \nn \,.
\eeqa
Expanding again as in (\ref{pseudoansatz}) and using again the integrals~(\ref{integrals}) leads to an expansion 
\beq
E_0= R{-}\frac{\pi}{3 R}{-}\frac{\pi^2}{18 R^3}{-}\frac{\pi^3}{54 R^5}{-}\frac{5\pi^4}{648 R^7}-\frac{128\pi^6 \alpha_3}{225 R^7}{+}\dots \la{finalE4D}  
\eeq
perfectly reproducing~\eqref{spec} and \eqref{nonunivbounds4D}.

\subsection{Level Splitting} \la{LevelSplitting}
Another quantity which is very sensitive to the deviation from universality of the S-matrix low energy parameters is the level splitting between various excited energy levels of the flux tube. In three dimensions, for example, their degeneracy is broken precisely by the non-universal parameter $\gamma_3$ so we can translate the bound $\gamma_3>-1/768$ directly into a bound on how much the levels can split. This is simplest to do in the integrability context where those energy levels are given by a TBA generalization valid for the excited states~\cite{Dorey:1996re}. 

Here they simplify quite a lot due to the bosonic nature of the particles (very rare in integrable models) which allows for particles to have the same momenta and also because of the absence of $LL$ and $RR$ scattering in this spontaneous symmetry breaking setup. If we consider $N$ right movers with the same mode number~$n$ and $N$ left movers with the same mode number~$-n$ then each quanta will have the same momentum~$+p/-p$ for right/left movers respectively where $p$ is quantized through a souped up set of Bethe ansatz equations which include finite size corrections and which also yield the~($D=3$) exited state energy as 
\beqa
&E&=R+2 N p+\int\limits_0^\infty\frac{dx}{\pi R}\log(1-e^{-\varepsilon(x)}) \!\!\! \\
&2\pi n&=p R{+}2 N\delta\left(\frac{4p^2}{R^2}\right){+}\!\!\int\limits_0^\infty \frac{dx}{2\pi x} \mathcal{K}\left(\frac{4p x }{R^2}\right) \log(1{-}e^{{-}\varepsilon(x)}) \,, \nn \\
&\varepsilon(x)&{=}x{+}\,2 N\delta\left(\frac{4x p}{R^2}\right){+}\int\limits_0^\infty \frac{dx^\prime}{2 \pi x^\prime}\mathcal{K}\left(\frac{4x x^\prime}{R^2}\right)\log(1{-}e^{{-}\varepsilon(x^\prime)})\,,\nonumber 
\eeqa
which can be solved using the ansatz in~\eqref{pseudoansatz} for the pseudo-energy and an ansatz of the form
\beq
p=\frac{2\pi n}{R}+\frac{p_1}{R^3}+\frac{p_2}{R^5}+\frac{p_3}{R^7}+\mathcal{O}\left(\frac{1}{R^8}\right).
\eeq
for the momentum. Note, without the need for any computation, that if the phase shift is linear $\delta(x) \sim x$, as it is for the first few universal terms, then we can rescale $p$ and the corresponding mode number to absorb $N$ completely leading to the above mentioned degeneracy: the energy only depends on $N\times n$ in this case. The breaking of the degeneracy will thus be directly proportional to the first non-universal deviation from the $e^{is}$ S-matrix. The simplest example is (\ref{splitting42}) which upon using our bound implies the upper bound 
\beq
E_{N=2,n=2}-E_{N=4,n=1}\leq \frac{9592\,\pi^6}{15 R^7}+O(R^{-9}) \la{finSplitting} \, , 
\eeq
on the degeneracy of the first two degenerate states. 

While a small degeneracy is indeed nicely measured in the lattice, it is unfortunately quite challenging to compare it with this analytic bound. In short, because of all the $\pi$'s in the expressions above, we need a radius of about $R \sim 10 \,\ell_s$ to be able to properly order the terms in the low energy expansion while most lattice data is given up to $R\sim 6 \,\ell_s$. It would be very nice to produce larger radius data.

\section{Perturbative Flux Tube Computations} \la{perturbativeAppendix}

As explained in~\ref{ens} the strategy that we follow to compute~\rf{spec} consists in organizing the calculation by dividing the action \rf{lag} into two pieces 
\beq
A=A_\text{int}+A_{\cancel{int}} \, .  \label{actdev}
\eeq
$A_\text{int}$ produces an integrable $S$-matrix up to $O(s^3)$. This $S$-matrix is fed into the TBA, which returns the universal part of $E_0(R)$.
The piece $A_{\cancel{int}}$ consists of the leading order breaking of integrability. In particular, this invovles the leading non-universal operators -- see below for a clarification. 
We work at leading order in perturbation theory with the integrability breaking piece $A_{\cancel{int}}$.  
Thus,   the leading order non-universal contribution to the vacuum energy density is 
  \beq
\begin{minipage}[h]{0.12\linewidth}
        \vspace{0pt}
        \includegraphics[width=\linewidth]{./figs/K4.pdf}  
   \end{minipage}   =\frac{32\pi^6(2-D)((D{-}2)\alpha_3{+}(D{-}4)\beta_3)}{225 R^8}  \, ,
   \label{appdevb1}
      \eeq
      after regularizing out the zero mode.

Before closing this section we must explain two effects that we have glossed over. 
First, recall that the low energy universality of the NG theory implies that  all the  one-loop leading order scattering amplitudes are universal.
 In particular, the one-loop  $2\rightarrow 4$ processes are non-zero  away from $D=3,26$. 
Interestingly, this implies that, in the absence of further  massless degrees of freedom, the NG theory is not integrable
 away from $D=3,26$ \cite{Cooper:2014noa}. These one-loop six-point amplitudes can be reproduced with the following local operator  
\beq
 ( \partial_\mu \partial_\nu X^i )^2   \big[   (\partial_\rho X^j )^4   - \frac{1}{2} \partial_\rho X^j \partial_\sigma X^j \partial^\rho X^k \partial_\sigma X^k \big] \, ,  \label{op6}
\eeq 
where   the overall normalization is unimportant for our purposes. 
Generically, \rf{op6} implies $O(1/R^7)$ deviations from the finite volume  spectrum associated to  the integrable $S$-matrix in appendix \ref{appendixtba}. Indeed, \rf{op6} is added in $A_\text{int}$ and subtracted in $A_{\cancel{int}}$. For the vacuum energy density, we must compute a single insertion of \rf{op6}
  \beq  \label{nonumber}
\begin{minipage}[h]{0.11\linewidth}
        \vspace{0pt}
        \includegraphics[width=\linewidth]{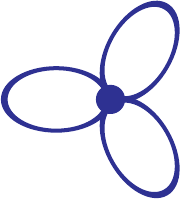}  
   \end{minipage}   =    \partial_\nu \partial_\alpha \partial_\beta \Delta_R(0) \,  \partial_\nu \partial_\gamma \partial_\beta \Delta_R(0) \, \partial_\alpha \partial_\gamma \Delta_R(0)  =0  \, , \nonumber
   \eeq
   where we omitted the symmetry factor. 
   Given that the latter  diagram vanishes, we conclude that at $O(1/R^7)$ the vacuum energy is insensitive to \rf{op6} and therefore the unique deviation from the square root formula at $O(1/R^7)$ is   given by \rf{appdevb1}.
   
In $D=3$,   the presence of the leading  non-universal operator induces  $2\rightarrow 4$  particle production at~$O(\ell_s^8 s^5)$  in the  $M$-matrix element \cite{Chen:2018keo}.  
Therefore, one could expect an $O(1/R^9)$ contribution to  the vacuum energy from this process. 
However, similarly to \rf{op6}, this process can be reproduced by a local operator $(\partial_\mu \partial_\nu X)^2 (\partial_\rho X)^2$ that has a vanishing  vacuum expectation value at finite volume.  

\medskip

The second effect is about the Einstein-Hilbert operator ${\cal R}$ introduced in \ref{sec:introduction}.
It is an evanescent operator, i.e. its contribution to the tree-level $S$-matrix vanishes for $d=2$ world-sheet spacetime dimensions; but, when dressed with virtual corrections in dimensional regularization,  gives a  non-zero contribution to the  $S$-matrix.
Its presence is needed for a consistent one-loop renormalization of the NG theory. Thus  ${\cal R}$ appears as a~$1/\epsilon+\om_3$ counter-term in dimensional regularization \cite{Dubovsky:2012sh}, where $\om_3$ is a finite non-universal choice for the counter-term.  
At~$O(s^3)$ the $2\rightarrow 2$ $S$-matrix of the renormalized NG theory involves two-loop Feynman diagrams from the NG vertices~$\sqrt{-\text{det}\, \partial_\alpha X^\mu \partial_\beta X_\mu}$ and one loop diagrams with a single insertion of the ${\cal R}$ and NG vertices. 
A priori, one could expect that~$\om_3$ together with the constants $\alpha_3$ and $\beta_3$ in \rf{lagalphbet} make up a triad of possible non-universal deformations  at $O(s^3)$.
However, the~$O(s^3)$ contribution of $\om_3$ is  analytic and thus can be absorbed in the~$K^4=O(\partial^8 X^4)$ operators, i.e. in a shift of the~$\{\alpha_3, \beta_3\}$ parameters \cite{Conkey:2016qju}.


\section{Axionic String Ansatz: no resonances}
\label{asa}
Here we elaborate further on the analysis of \ref{ftd3}, by  assuming  that there are no resonances on the $D=3$ flux tube, namely the Axionic String Ansatz (ASA)~\cite{Dubovsky:2015zey}.
If there are no resonances,  then the phase shift $\delta(s)= \frac{1}{2i} \log S(s)$ is analytic in the upper half plane. 
Let us derive a dispersion relation for the phase shift. We start from the identity
\beq
\frac{2\delta(s)-\frac{s}{4}}{s^2}   = \oint_s \frac{dz}{2\pi i} \left[
\frac{1}{z-s} -\frac{1}{z+s}\right] \frac{2\delta(z)-\frac{z}{4}}{z^2}  
\eeq
where the contour goes around $s$ in the upper half plane.
Assuming that $\delta(s)/s^3 \to 0 $ for $|s| \to \infty$ in the upper half plane, we can open the contour to the real axis to obtain
\beq
 2\delta(s)    =  \frac{s}{4}+ \frac{2 s^3}{\pi} \int_{-\infty}^\infty dz  
\frac{{\rm Im}\, \delta (z) }{z^2(z^2-s^2)}  
\eeq
This gives
\beq
\gamma_3/\gamma_7= \langle z^4 \rangle \,,\quad
\gamma_5/\gamma_7= \langle z^2 \rangle 
\eeq
with
\beq
\langle z^n \rangle = \int_{-\infty}^\infty dz \, \rho(z) z^n\,,\qquad
\rho(z)= \frac{2 }{\pi \gamma_7 } 
\frac{{\rm Im}\, \delta (z) }{z^8} \,.
\eeq
Notice that $\rho(z)=\rho(-z)$ is a non-negative normalized distribution $\int  dz \,\rho(z)=1$.
Therefore, we conclude the 
\beq
\gamma_3,\gamma_5,\gamma_7 \ge 0\,.
\label{ASAposconst}
\eeq
Furthermore the large $s$ behavior of the phase shift is given by
\beq
2\delta(s)    = \left( \frac{1}{4}-  \gamma_7\langle z^6 \rangle  \right) s +\dots
\eeq
Causality allows for a time-delay but not a time advance. 
This implies that the coefficient of the linear term in $s$ at high energies must be positive \cite{Dubovsky:2012wk,Camanho:2014apa}.
In other words, 
\beq
\ell_{UV}^2=\ell_s^2\left(1-4  \gamma_7 \langle z^6 \rangle \right)>0\,.
\eeq

We can derive more inequalities. Firstly,
\beq
\left\langle \left(z^2 -\langle z^2 \rangle \right)^2 \right\rangle \ge 0 \Rightarrow
\gamma_3 \gamma_7 \ge \gamma_5^2
\eeq
Secondly,
\beq
\left\langle z^2\left(z^2 -\langle z^2 \rangle \right)^2 \right\rangle \ge 0 \Rightarrow
\langle z^6 \rangle\ge 2\langle z^4 \rangle\langle z^2 \rangle-\langle z^2 \rangle^3
\eeq
together with $\langle z^6 \rangle\le \frac{1}{4\gamma_7}$, leads to
\beq
 \gamma_7^2 \ge 4 \gamma_5(2\gamma_3\gamma_7-\gamma_5^2)\,.
\eeq
These give an upper and lower bound on $\gamma_3$,
\beq
\label{ASAgamma3}
\frac{ \gamma_5^2}{ \gamma_7} \le \gamma_3  \le \frac{  \gamma_7^2 +4  \gamma_5^3}{ 8 \gamma_5 \gamma_7}\,.
\eeq
This  must fit inside the  allowed region in figure \ref{3Dfigure}.
Notice that the new region has an edge where the upper and lower bounds coincide,
\beq
 \gamma_7^2 = 4\gamma_5^3 =4 \gamma_3 \gamma_5 \gamma_7\,.
\eeq
At this edge, $\ell_{UV}=0$ because
\beq
 \int_{-\infty}^\infty dz  
\frac{{\rm Im}\, \delta (z) }{z^2 }  = \frac{\pi}{8}
\eeq
and the distribution is delta-function type, 
$$\rho(z)=\frac{1}{2}\delta(z-z_0)+\frac{1}{2}\delta(z+z_0)
$$ 
where $4\gamma_7 z_0^6=1$.

Can we confirm or disprove the ASA?
In principle, the ASA can be disproved if one measures low energy constants $\gamma_3,  \gamma_5, \dots $ incompatible with the constraints
\eqref{ASAposconst} or \eqref{ASAgamma3}.
Currently, the available estimate $\gamma_3 \approx 3\times 10^{-4}$ (from lattice data for $SU(6)$ YM~\cite{Athenodorou:2011rx,Dubovsky:2014fma, Chen:2018keo}) is compatible with the ASA.

\bibliography{thefluxtube} 

\end{document}